\documentclass[journal]{IEEEtran}

\usepackage{cite}

\ifCLASSINFOpdf
   \usepackage[pdftex]{graphicx}
\else
   \usepackage[dvips]{graphicx}
\fi

\usepackage[caption=false]{subfig}
\usepackage{algorithmic}
\usepackage[ruled]{algorithm2e}

\usepackage{manfnt}
\makeatletter
    \def\@endtheorem{\hspace*{0pt}\hfill$\square$\@endpefalse }
\makeatother

\usepackage{epsfig}
\usepackage{epstopdf}

\usepackage{graphics} 
\usepackage{epsfig} 
\usepackage[cmex10]{amsmath} 
\usepackage{amssymb}  
\usepackage{microtype}
\usepackage{bm}
\usepackage{color}
\usepackage[dvipsnames]{xcolor}
\usepackage{tikz}
\usepackage{microtype}
\usetikzlibrary{shapes,arrows}
\usepackage{url}
\usepackage{hyperref}

\DeclareMathOperator{\sign}{sign}

\DeclareMathOperator{\diag}{diag}
\DeclareMathOperator{\rank}{rank}

\DeclareMathOperator{\R}{\mathbb{R}}
\DeclareMathOperator{\cR}{\mathcal{R}}

\DeclareMathOperator{\W}{\mathcal{W}}
\DeclareMathOperator{\X}{\mathcal{X}}
\DeclareMathOperator{\p}{\mathcal{P}}
\DeclareMathOperator{\Sys}{\mathcal{S}}
\newcommand{\vect}[1]{\text{vec}\left(#1\right)}
\newcommand{\red}[1]{\widetilde{#1}}
\everymath{\displaystyle}

\newtheorem{theorem}{Theorem}
\newtheorem{proposition}{Proposition}
\newtheorem{corollary}{Corollary}

\newtheorem{lemma}{Lemma}
\newtheorem{remark}{Remark}
\newtheorem{problem}{Problem}
\newtheorem{example}{Example}
\newtheorem{assumption}{Assumption}
\newtheorem{definition}{Definition}

\title{On Moment Matching for Stochastic Systems}

\author{Giordano Scarciotti and Andrew R. Teel
\thanks{G. Scarciotti is with the Dept. of Electrical and Electronic Engineering, Imperial College London, London, SW7 2AZ, UK, E-mail: g.scarciotti@ic.ac.uk.}
\thanks{A.R. Teel is with the Electrical and Computer Engineering Department, University of California, Santa Barbara, CA 93106-9560 USA, E-mail: teel@ece.ucsb.edu)}
\thanks{This work was supported in part by Imperial College London under the Junior Research Fellowship Scheme, and in part by NSF grant no. ECCS-1508757 and by AFOSR grants AFOSR FA9550-15-1-0155 and FA9550-18-1-0246.}%
}%

\begin{document}

This article has been accepted for publication by IEEE Transactions on Automatic Control.

\vspace{1cm}

The manuscript included in this file is the open access accepted version. 

\vspace{1cm}

This open access version is released on arXiv in accordance with the IEEE copyright agreement.

\vspace{1cm}

The final version will be available at (not open access) \url{https://doi.org/10.1109/TAC.2021.3050711}

\maketitle

\begin{abstract}
In this paper we study the problem of model reduction by moment matching for stochastic systems. We characterize the mathematical object which generalizes the notion of moment to stochastic differential equations and we find a class of models which achieve moment matching. However, differently from the deterministic case, these reduced-order models cannot be considered ``simpler'' because of the high computational cost paid to determine the moment. To overcome this difficulty, we relax the moment matching problem in two different ways and we present two classes of reduced-order models which, \textit{approximately} matching the stochastic moment, are computationally tractable.
\end{abstract}


\section{Introduction}
\label{MRSto-sec1}

\IEEEPARstart{D}{ynamical} systems described by stochastic differential equations have been successfully used in a variety of theoretical and applied scientific fields, such as system biology and finance \cite{KarShr:91,YonZho:99,Mun:11,Oks:13} (see also the seminal contributions of Kalman \cite{Kal:60,KalBuc:61} in control theory). One way to view these systems is to interpret the stochastic processes (\textit{i.e.} time sequences representing the evolution of variables which are subject to random variations) in the equations as a means to model \textit{uncertainty}. In this sense, stochastic systems offer a powerful modeling framework for engineering applications. However, as noted in \textit{e.g.} \cite{LinPic:15}, the complexity of the algorithms solving stochastic problems usually grows more than linearly with the dimension of the model. For this reason, many researchers have investigated the problem of model reduction for stochastic systems. The objective of model reduction is to obtain a \textit{simple}, in some sense to be defined, model of the original system which possesses, for some operating conditions, a specific subset of the properties of the system to be reduced. Most of this field originally developed from the need of reducing the order of high-order linear ordinary differential equations obtained from the discretisation of partial differential equations for simulation purposes, see \cite{Ant:05}. From there, the field expanded to much more general classes of systems, such as time-delay and nonlinear systems \cite{ScaAst:17}, and purposes, such as analysis and control, see for instance \cite{FaeScaAstRin:18,BreForScaAst:19,FaeScaAstRin:21}. Many stochastic generalizations of deterministic methods have been proposed. For instance for linear systems various forms of stochastic balancing truncation have been presented in \textit{e.g.}\cite{DesPal:84,DesPalKir:85,Tug:85,Gre:88,Gre:88a,HarJonSil:84,WanSaf:90,WanSaf:91,LinPic:96}, whereas stochastic $\mathcal{H}_{\infty}$ model reduction has been given in \textit{e.g.} \cite{XuChe:03,SuWuShiSon:12}. For bilinear systems, balanced truncation for stochastic systems has been proposed in \cite{BenDam:11,BenRed:15}. For nonlinear systems, the slaving principle, the slow-fast reduction based on center manifold theory,  and the use of eigenfunctions are all methods which have been extended to stochastic systems, see \cite{SchHak:86,SchHak:87}, \cite{XuRob:96,Rob:08,Rob:14} and \cite{CoiKevLafMagNad:08}. Moreover, model reduction of stochastic systems continues to be an active area of research for applications such as quantum stochastic systems, see \textit{e.g.} \cite{Nur:14,TecNur:16}, and biological and chemical systems, see \textit{e.g.} \cite{PelMunKha:06,SinHes:10,BruChaSmi:14,SooAnd:14,GupKha:14,MelHesKha:14,JohBarSejMjo:15,SmiCiaGri:15}.
However, other successful deterministic approaches for model reduction of nonlinear systems, such as \cite{Sch:93} (nonlinear balancing) and \cite{Ast:10} (nonlinear moment matching), have not yet been extended to the stochastic framework.

In this paper we address the problem of model reduction for a general class of linear and nonlinear stochastic systems using a moment matching approach. By ``general class'', we mean systems in which the state and the input may appear simultaneously in both the drift term and diffusion term of the equation, \textit{i.e.} they may multiply both the $dt$ term and the $d\W_t$ term, where $\W_t$ is a Brownian motion. Borrowing from the deterministic literature, we formulate the problem of model reduction by moment matching as the problem of determining a reduced-order model that possesses the same steady-state output of the system to be reduced for specific classes of inputs. For the sake of studying the problem in its most general form, also the signal generator which produces the input signals of interest is selected as a stochastic system. To the end of matching the steady-state output of the system, we construct a stochastic process, which we call moment, that describes the steady-state behaviour of the system. The moment is the solution of a stochastic generalization of the Sylvester equation for linear systems and of the invariance partial differential equation for nonlinear systems. We then propose families of reduced-order models that match the moment of the system to be reduced. Unfortunately, it is shown that the reduced-order models have an on-line computational complexity which is comparable to the one of the original system. To overcome this issue two strategies are adopted to provide families of approximated models. The first family preserves a part of the stochastic properties of the moment, whereas the second family preserves a part of the stochastic properties of the steady-state output. Finally, the differences between these families of reduced-order models are illustrated by means of simulations of numerical examples and of a stochastic variation of a deterministic benchmark system given in \cite{ChaVaD:05,SLICOT}.

Preliminary versions of our work have been published in \cite{ScaTee:17,ScaTee:17a}. Note that considering a stochastic signal generator is a nontrivial generalisation of \cite{ScaTee:17,ScaTee:17a}. First, many of the assumptions have to be revised and relaxed. Second, a stochastic signal generator has the peculiarity of changing the mean of the mappings describing the steady-state response, which is a feature that is not captured in \cite{ScaTee:17,ScaTee:17a}. In this regard, a new family of models (moment-mean) is proposed and several variations of the results are sketched (for instance considering multiple Brownian motions). Third, the complete proofs of the results are provided. These are of independent interest beyond the model reduction literature because many of the results are instrumental to the solution of other mathematical control problems, such as output regulation \cite{Isi:95,Sca:18,MelSca:19,MelSca:19a,MelSca:20}. Finally, new simulations, numerical and based on a benchmark example, are provided.

The rest of the paper is organized as follows. In Section~\ref{MRSS-sec-PN} we formulate the problem. In Section~\ref{MRSto-SecSS} we characterize the moment for linear and nonlinear stochastic systems. In Section~\ref{MRNS-secROM} we provide families of reduced-order models which match the moment of the system to be reduced. In Section~\ref{MRSS-sec-AROM} we first point out that the found models have a high computational complexity and then we propose new classes of approximated reduced-order models. In Section~\ref{MRSS-sec-Sim} we illustrate some of the results by means of simulations. Section~\ref{MRSS-sec-Con} contains our concluding remarks. In order to make the paper as self-contained as possible, the most important concepts of stochastic systems used throughout the paper are reported, with plenty of references, in the Appendix.

\textbf{Notation.} We use standard notation. $\mathbb{C}_{<0}$ ($\mathbb{C}_{\ge 0}$) denotes the set of complex numbers with negative (non-negative) real part. $\mathbb{R}_{<0}$ ($\mathbb{R}_{> 0}$) denotes the set of negative (positive) real numbers. The symbol $I$ denotes the identity matrix and $\sigma(A)$ denotes the spectrum of the matrix $A\in\mathbb{R}^{n\times n}$. The symbols $|v|$, with $v\in\R^n$, and $||A||$ indicates the Euclidean norm and the induced Euclidean matrix norm, respectively. The vectorization of a matrix $A\in\mathbb{R}^{n\times m}$, denoted by $\vect{A}$, is the $nm \times 1$ vector obtained by stacking the columns of the matrix $A$ one on top of the other, namely $\vect{A}=[a_1^{\top},a_2^{\top},\dots,a_m^{\top}]^{\top}$, where $a_i\in\mathbb{R}^n$ is the $i$-th column of $A$ and the superscript $\top$ denotes the transposition operator. The symbol $\otimes$ indicates the Kronecker product, whereas the symbol $\oplus$ indicates the direct sum. Given two functions, $f:Y\to Z$ and $g:X\to Y$, with $f\,\circ\,g:X\to Z$ we denote the composite function $(f\,\circ\,g)(x)=f(g(x))$ which maps all $x\in X$ to $f(g(x))\in Z$. $(\Omega,\mathcal{F},\mathcal{P})$ indicates a probability space with a given set $\Omega$, a $\sigma$-algebra $\mathcal{F}$ on $\Omega$ and a probability measure $\mathcal{P}$ on the measurable space $(\Omega,\mathcal{F})$. For ease of notation, we often indicate a stochastic process $\{x_t,\,t\in\mathbb{R}\}$ simply with $x_t$ (this is common in the literature, see \textit{e.g.} \cite{Arn:74}).  
The stochastic process $\W_t$ indicates a standard Wiener process defined on the probability space $(\Omega,\mathcal{F},\mathcal{P})$. $\mathcal{F}_t$ is the continuous-time filtration generated by the Wiener process $\W_t$ up to time $t$ and all stochastic processes appearing in this paper are adapted to this filtration (and, possibly, to others when multiple Brownian motions appear). Let $X_t$ and $Y_t$ be any two stochastic processes and $d: (t,x,y) \mapsto d_t(x,y) $ any function which has well-defined partial derivatives; then any partial derivative of the form 
$\frac{\partial d_t}{\partial x} (X_t,Y_t)$ is compactly indicated as $\frac{\partial d_t}{\partial X_t}$.
All the stochastic integrals in this paper are intended as It\^{o} integrals.

\section{Problem formulation}
\label{MRSS-sec-PN}

In this section we formulate the problem of model reduction by moment matching for stochastic systems. Complete definitions of stochastic process, Brownian motion, It\^{o}'s formula, stability and so on are reported in the Appendix. The reader who is not familiar with stochastic differential equations is invited to consult the Appendix and references therein.

Consider a stochastic nonlinear single-input single-output continuous-time system $\Sys_{n}(f,g,h)$ described by the equations
\begin{equation}
\label{MRNS-eq1}
dx_t=f(x_t,u_t)dt+g(x_t,u_t)d\W_t, \qquad y_t=h(x_t),
\end{equation}
with $x_t\in\mathbb{R}^n$, $u_t\in \mathbb{R}$, $y_t\in\mathbb{R}$, and $f$, $g$ and $h$ smooth mappings. 
Consider a signal generator $\Sys_{\nu}(s,j,l)$ described by the equations
\begin{equation}
\label{MRNS-eq2}
d \omega_t=s(\omega_t)dt + j(\omega_t)d\W_t , \qquad u_t=l(\omega_t),
\end{equation}
with $\omega_t\in \mathbb{R}^{\nu}$, and $s$, $j$ and $l$ smooth mappings. 
Consider the interconnection of system~(\ref{MRNS-eq1}) with the signal generator~(\ref{MRNS-eq2}), namely
\begin{equation}
\label{MRNS-eqInt}
\!\!\!\!\!\begin{array}{l}
\left[\!\!\!\begin{array}{c}d \omega_t \\ dx_t \end{array}\!\!\!\right]\!=\!\left[\!\!\!\begin{array}{c} s(\omega_t) \\ f(x_t,l(\omega_t)) \end{array}\!\!\!\right] \!dt \!+\! \left[\!\!\!\begin{array}{c}j(\omega_t) \\ g(x_t,l(\omega_t))\end{array}\!\!\!\right]\!d\W_t, \quad y_t=h(x_t).
\end{array}
\end{equation}
Assume that zero is an equilibrium point of (\ref{MRNS-eqInt}), \textit{i.e.} $s(0)=0$, $j(0)=0$, $l(0)=0$, $f(0,0)=0$, $g(0,0)=0$ and $h(0)=0$, and that the initial condition $(\omega(0),x(0))$ is deterministic. 



We now define the ``moment matching condition'' and introduce the ``problem of model reduction by moment matching''. An intuitive explanation follows the problem.

\begin{definition}
Consider system~(\ref{MRNS-eq1}) and the signal generator~(\ref{MRNS-eq2}). A stochastic system $\Sys_{\nu}(\red{f},\red{g},\red{h})$ described by the equations
\begin{equation}
\label{MRSto-eqRM}
d\red{x}_t=\red{f}(\red{x}_t,u_t)dt+\red{g}(\red{x}_t,u_t)d\W_t, \qquad \red{y}_t=\red{h}(\red{x}_t),
\end{equation}
where $\red{x}_t\in\mathbb{R}^{\nu}$, with $\nu < n$, $\red{y}_t\in\mathbb{R}$, and $\red{f}$, $\red{g}$ and $\red{h}$ are smooth mappings, is said to satisfy the \textit{moment matching condition at $(s,j,l)$} if the error $e_t=y_t-\red{y}_t$, where $y_t$ is the output of (\ref{MRNS-eqInt}) and $\red{y}_t$ is the output of (\ref{MRSto-eqRM}) driven by (\ref{MRNS-eq2}), satisfies
\begin{equation}
\label{MRSto-eqMMc}
\lim_{t\to \infty} e_t = 0
\end{equation}
almost surely, for any $(x_0,\omega_0,\red{x}_0)\in \mathcal{N} \subset \mathbb{R}^{n} \times \mathbb{R}^{\nu} \times \mathbb{R}^{\nu}$.
\end{definition}

\begin{problem} 
\label{MRSto-pro}
Consider system~(\ref{MRNS-eq1}) and the signal generator~(\ref{MRNS-eq2}). The \textit{problem of model reduction by moment matching} consists in determining a stochastic system $\Sys_{\nu}(\red{f},\red{g},\red{h})$, with $\nu<n$, which satisfies the moment matching condition at $(s,j,l)$.
\end{problem}

\begin{figure} {\color{black}
\centering 
\tikzset{%
  block/.style    = {draw, thick, rectangle, minimum height = 3em,
    minimum width = 3em,text width=5.5em, text centered},
  sum/.style      = {draw, circle, node distance = 2cm}, 
  dot/.style      = {coordinate}, 
  input/.style    = {coordinate}, 
  output/.style   = {coordinate} 
}
\begin{tikzpicture}[auto, thick, node distance=2cm, >=triangle 45]
\draw
	node at (0,-1)[block] (input1) {$\Sys_{\nu}(s,j,l)$} 
	node at (1.7,-1)[dot] (suma1) {}
	node at (3.5,0)[block] (inte1) {$\Sys_{n}(f,g,h)$}
    node at (5.3,-1)[sum] (suma2) {$-$}
    node at (3.5,-2.0)[block] (ret1) {$\Sys_{\nu}(\red{f},\red{g},\red{h})$}
	node at (7.4,-1) [output] (sal2){}; 
    \draw[-](input1) -- node {$u_t$} (suma1);
    \draw[->](suma1) |- node {} (inte1);
 	\draw[->](suma1) |- node {} (inte1);
	\draw[->](inte1) -| node [near end, left] {$y_t$} (suma2);
	\draw[->](ret1) -| node [near end, left] {$\red{y}_t$} (suma2);
	\draw[->](suma1) |- node {} (ret1);
	\draw[->] (suma2) -- node {$ e_t \to 0$ a.s.}(sal2);
\end{tikzpicture} }
\caption{A schematic overview of Problem~\ref{MRSto-pro}: given $\Sys_{\nu}(s,j,l)$ and $\Sys_{n}(f,g,h)$ we want to determine $\Sys_{\nu}(\red{f},\red{g},\red{h})$ such that $ e_t \to 0$ almost surely.} 
\label{MRDI-fig1}
\end{figure}
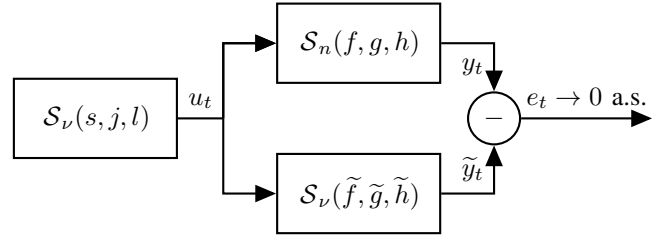

The interpretation of Problem~\ref{MRSto-pro}, as illustrated in Fig.~\ref{MRDI-fig1}, is that we are looking for a family of \textit{simpler} models (where simplicity is somewhat arbitrarily intended as $\nu<n$) which behave asymptotically as the system that we want to reduce when both the system and the reduced-order model are driven by an \textit{a priori} selected signal generator. In other words, we are interested in preserving the asymptotic behavior of the system for specific operating conditions and input signals. This ``steady-state matching'' method is called ``moment matching'' because \cite{Ast:10} recognized that for deterministic linear systems the method is equivalent to matching the ``moments'' (\textit{i.e.} the coefficients of a series expansion of the transfer function) as defined in \cite{Ant:05}. We anticipate from the onset that we also solve ``approximated'' versions of Problem~\ref{MRSto-pro} in which the moment matching condition~(\ref{MRSto-eqMMc}) is relaxed in various ways.


At this point, it is useful to introduce the notation for stochastic linear systems because, in addition to providing stronger and more easily interpretable results, this class of systems has a large practical use in the deterministic model reduction literature. Hence, we expect that the linear results of this paper will have a more immediate impact. When system~(\ref{MRNS-eq1}) is linear, we use the notation
\begin{equation}
\label{ORS-eq1}
dx_t=(Ax_t+Bu_t)dt+(Fx_t+Gu_t)d\W_t,\qquad y_t=Cx_t,
\end{equation}
with $A\in \mathbb{R}^{n\times n}$, $B\in \mathbb{R}^{n\times1}$, $F\in \mathbb{R}^{n\times n}$, $G\in \mathbb{R}^{n\times1}$ and $C\in \mathbb{R}^{1\times n}$. Similarly, when the generator~(\ref{MRNS-eq2}) is linear, we use the notation 
\begin{equation}
\label{ORS-eq2}
d\omega_t= S\omega_t dt+J\omega_t d\W_t, \qquad u_t=L\omega_t,
\end{equation}
with $S\in \mathbb{R}^{\nu\times \nu}$, $J\in \mathbb{R}^{\nu\times\nu}$ and $L\in \mathbb{R}^{1\times \nu}$. 
Let $\Phi_t\in\mathbb{R}^{n\times n}$ be the fundamental matrix of the homogeneous equation corresponding to (\ref{ORS-eq1}), \textit{i.e.}
\begin{equation}
\label{ORS-eqPhi}
d\Phi_t =  \left( A  dt + F d\W_t\right) \Phi_t,
\end{equation}
with $\Phi_0=I$ and recall that (see \cite[Section 4.1]{Gar:88}) 
\begin{equation}
\label{ORS-eqPhiInv}
d\Phi_t^{-1} = \Phi_t^{-1} \left( (F^2-A) dt - F d\W_t \right).
\end{equation}
Finally, let $\Sigma_t\in\mathbb{R}^{\nu\times \nu}$ be the fundamental matrix corresponding to (\ref{ORS-eq2}), \textit{i.e.}
\begin{equation}
\label{ORS-eqSigma}
d\Sigma_t =  \left( S  dt + J  d\W_t \right) \Sigma_t ,
\end{equation}
holds.

\section{Steady state of stochastic systems}
\label{MRSto-SecSS}

Since the moment matching condition~(\ref{MRSto-eqMMc}) is a condition on the steady-state behavior of the system, it is instrumental for the solution of the problem to provide a description of the steady-state response of the system. In this section we characterize the steady state of system~(\ref{ORS-eq1}) driven by (\ref{ORS-eq2}) in terms of a stochastic partial differential equation, and of system~(\ref{ORS-eq1}) driven by (\ref{ORS-eq2}) in terms of a stochastic Sylvester equation. Note that these two results, which are the main technical contributions of the paper, have an interest beyond the problem of model reduction (because, for instance, similar characterizations of the steady-state response are an essential element of the solution of output regulation problems \cite{Isi:95,Sca:18}).

We begin with the linear case, which is easier to develop and is instrumental for the nonlinear case. At the end of the section we  show that the results can be formulated also for the simpler case of  multiple uncorrelated Brownian motions acting on the system and/or the signal generator.

\subsection{Steady state: linear stochastic systems}
\label{MRSto-SubLSS}

We introduce the following assumptions\footnote{For the definition of Lyapunov exponent and a procedure to check these assumptions, we refer the reader to the Appendix.}.
\begin{assumption}
\label{ORS-asSys}
All Lyapunov exponents of $\Phi_t$ are negative almost surely.
\end{assumption}
\begin{assumption}
\label{ORS-asGen}
All Lyapunov exponents of $\Sigma_t$ are zero almost surely.
\end{assumption}

We are now ready to give a characterization of the steady-state response of system~(\ref{ORS-eq1}) driven by~(\ref{ORS-eq2}).

\begin{theorem}
\label{ORS-thLSSX}
Consider the interconnection of system~(\ref{ORS-eq1}) and the signal generator~(\ref{ORS-eq2}). Suppose that Assumptions~\ref{ORS-asSys} and \ref{ORS-asGen} hold. Then the steady-state response of the output of such interconnection is
$$
y^{ss}_t=C \X_t\omega_t
$$
almost surely, where $\X_t\in\mathbb{R}^{n\times \nu}$ is
\begin{equation}
\label{ORS-eq4}
\!\!\X_t\!=\!\Phi_t \!\left[\int_{-\infty}^t\!\!\!\!\!\! \Phi_{\tau}^{-1}(BL -FGL)\Sigma_{\tau}d\tau\!+\!\!\!\int_{-\infty}^t \!\!\!\!\!\! \Phi_{\tau}^{-1}GL\Sigma_{\tau}d\W_{\tau}\right]\!\Sigma_{t}^{-1}\!\!.
\end{equation}
The stochastic process $\X_t$ is the steady-state solution of the stochastic differential matrix equation
\begin{equation}
\label{ORS-eqPI}
\begin{array}{rl}
d\X_t=\!\!\!\!&\left(A\X_t-\X_t\left(S-J^2\right)-F\X_t J+BL-GLJ\right)dt\\[2mm]
&+\left(F\X_t-\X_t J+GL\right)d\W_t.
\end{array}
\end{equation}
\end{theorem}

\textit{Proof:}
Consider the matrix $\X_t$ defined in (\ref{ORS-eq4}). Multiplying this equation by $\Phi_t^{-1}$ on the left and by $\Sigma_t$ on the right yields
\begin{equation}
\label{MRLS-eqSSp1}
\begin{array}{l}
\!\!\!\!\!\!\!\!\!\!\!\!\!\!\!\int_{-\infty}^t  \!\!\!\!\!\! \Phi_{\tau}^{-1}(BL-FGL) \Sigma_{\tau} d\tau + \int_{-\infty}^t  \!\!\!\!\!\! \Phi_{\tau}^{-1}GL \Sigma_{\tau}d\W_{\tau},\\
=\Phi_t^{-1}\X_t \Sigma_t=\int_{-\infty}^t d\!\left(\Phi_{\tau}^{-1} \X_{\tau} \Sigma_{\tau} \right),
\end{array}
\end{equation}
in which the first equality holds because of (\ref{ORS-eq4}) and the second equality holds because $\textstyle\lim_{t\to-\infty}\Phi_t^{-1}\X_t \Sigma_t=0$ almost surely. In fact, by Assumptions~\ref{ORS-asSys} and \ref{ORS-asGen}, $\textstyle\lim_{t\to-\infty}\Phi_t^{-1}=0$ exponentially, $\textstyle\lim_{t\to-\infty}\frac{1}{t}\Sigma_t=0$ and $\textstyle\lim_{t\to-\infty}\frac{1}{t}\X_t=0$. This last limit holds for the same reasons noting that (\ref{ORS-eq4}) multiplied by $\omega_t$ is a response of (\ref{ORS-eq1}) to the input (\ref{ORS-eq2}). We can now use Leibniz integral rule to differentiate the integrals in both sides of~(\ref{MRLS-eqSSp1}) obtaining
$$
\begin{array}{l}
\Phi_t^{-1}(BL-FGL)\Sigma_tdt+\Phi_t^{-1}GL\Sigma_td\W_t=d(\Phi_t^{-1}\X_t\Sigma_t)\\[2mm]
\,\,\,\,\,=d\Phi_t^{-1}\X_t \Sigma_t+\Phi_t^{-1}d\X_t \Sigma_t+\Phi_t^{-1}\X_t d\Sigma_t\\[2mm]
\,\,\,\,\,+d\Phi_t^{-1}d\X_t\Sigma_t+\Phi_t^{-1}d\X_t d\Sigma_t+d\Phi_t^{-1}\X_t d\Sigma_t,
\end{array}
$$
where the term $d\Phi_t^{-1}d\X_t d\Sigma_t$ does not appear because it is zero by~(\ref{ORS-eqIto1}) in the Appendix. Substituting (\ref{ORS-eqPhiInv}) and (\ref{ORS-eqSigma}) in the previous equation and multiplying on the left by $\Phi_t$ and on the right by $\Sigma_t^{-1}$, yields
$$
\begin{array}{rl}
\left((F^2\right.\!\!\!\!\!\!&-\left.A)dt-Fd\W_t\right)\X_t+d\X_t+\X_t (Sdt+Jd\W_t) \\[2mm]
&+\left((F^2-A)dt-Fd\W_t\right)d\X_t+d\X_t (Sdt+Jd\W_t)\\[2mm]
&+\left((F^2-A)dt-Fd\W_t\right)\X_t(Sdt+Jd\W_t)\\[2mm]
=\!\!\!\!&(BL-FGL)dt+GLd\W_t.
\end{array}
$$
By (\ref{ORS-eqIto1}) we note immediately that in the second line $dtd\X_t=0$ and in the third line only the term in $d\W_t d\W_t=dt$ is not zero. By sorting this expression we obtain
\begin{equation}
\label{MRLS-eqSSp2}
\begin{array}{rl}
\!\!d\X_t&\!\!\!=\left((A-F^2)\X_t-\X_t S + F\X_t J +BL-FGL\right)dt\\[2mm]
&\!\!\!+(F\X_t-\X_t J+GL)d\W_t + (Fd\X_t -d\X_t J)d\W_t.
\end{array}
\end{equation}
Multiplying (\ref{MRLS-eqSSp2}) first by $Fd\W_t$ on the left and then (separately) by $Jd\W_t$ on the right and using (\ref{ORS-eqIto1}) and (\ref{ORS-eqIto2}) in the Appendix, yields
$$
\begin{array}{rcl}
Fd\X_t d\W_t \!\!\!&=&\!\!\!(F^2\X_t-F\X_t J +FGL)dt,\\
-d\X_t J d\W_t \!\!\!&=&\!\!\! (-F\X_t J+\X_t J^2 -GLJ)dt.
\end{array}
$$
Substituting these two expressions in (\ref{MRLS-eqSSp2}) proves that $\X_t$, defined in (\ref{ORS-eq4}), is the solution of the stochastic differential matrix equation~(\ref{ORS-eqPI}). 
Now define the variable $z_t:=x_t-\X_t\omega_t$. Then by the stochastic product rule (recalled as Lemma~\ref{MRSS-lmdd} in the Appendix)
\begin{equation}
\label{MRSS-eqdz}
dz_t=dx_t-d\X_t\omega_t-\X_t d\omega_t - d\X_t d\omega_t.
\end{equation}
Note that $d\X_t d\omega_t = \left(F\X_t-\X_t J+GL\right)J\omega_t dt$
by (\ref{ORS-eqIto1}), (\ref{ORS-eqIto2}) and (\ref{ORS-eqPI}). Substituting the expressions of $dx_t$, namely (\ref{ORS-eq1}), of $d\X_t$, namely (\ref{ORS-eqPI}), and of $d\omega_t$, namely (\ref{ORS-eq2}), in (\ref{MRSS-eqdz}) yields
$$
\!dz_t\!=\!A(x_t-\X_t\omega_t)dt+F(x_t-\X_t\omega_t)d\W_t\!=\! Az_tdt+Fz_td\W_t\!.
$$
The zero equilibrium $z_t=0$ of this last equation is asymptotically stable almost surely because of Assumption~\ref{ORS-asSys}. In turn this proves that $x_t$ converges to $\X_t\omega_t$ almost surely as $t\to+\infty$. The claim follows substituting the steady-state response $x_{ss}=\X_t \omega_t$ in the equation of $y_t$ in (\ref{ORS-eq1}).
\hspace*{0pt}\hfill $\square$ \par

\subsection{Steady state: nonlinear stochastic systems}
\label{MRSto-SubNLSS}

We now generalize the previous result to nonlinear systems, \textit{i.e.} we show that the steady-state response of the interconnection of system~(\ref{MRNS-eq1}) and (\ref{MRNS-eq2}), namely (\ref{MRNS-eqInt}), can be characterized by a stochastic partial differential equation which is a generalization of \cite[Equation~(8.3)]{Isi:95}. To streamline the presentation we formulate the following assumptions.
\begin{assumption}
\label{MRNS-as1}
Assumption~\ref{ORS-asSys} holds for the linearization of system~(\ref{MRNS-eq1}) around the zero equilibrium.
\end{assumption}
\begin{assumption}
\label{MRNS-as2}
Assumption~\ref{ORS-asGen} holds for the linearization of system~(\ref{MRNS-eq2}) around the zero equilibrium.
\end{assumption}

These two assumptions induce a decomposition of the state space $\mathbb{R}^{n+\nu}$ as $\mathbb{R}^{n+\nu}=E_s(w)\oplus E_c(w)$, where $E_s$ and $E_c$ are  the stable and central Odeselec spaces, in a similar way as negative and zero real part eigenvalues induce a decomposition in stable and central subspaces for deterministic systems (see the Appendix for more details).
\begin{theorem}
\label{MRNS-thm1}
Consider system~(\ref{MRNS-eq1}) and the signal generator~(\ref{MRNS-eq2}). Suppose Assumptions~\ref{MRNS-as1} and \ref{MRNS-as2} hold. Then there exists a stochastic process $\chi:(t,\W_t,\omega_t)\mapsto \chi_t(\W_t,\omega_t)$, locally defined in a neighborhood $W(w) \subset E_c(w)$ of $\omega_t=0$, with $\chi_t(0,0)=0$, which solves the stochastic partial differential equation
\begin{equation}
\label{MRNS-eqNSS}
d \chi_t= f(\chi_t(\W_t, \omega_t),l(\omega_t))dt + g(\chi_t(\W_t, \omega_t),l(\omega_t))d\W_t,
\end{equation}
where
\begin{equation}
\label{MRNS-eqSPDE}
\begin{array}{rl}
\!\!\!d \chi_t &\!\!\!\!=\left[\frac{\partial \chi_t}{\partial t}+ \frac{\partial \chi_t}{\partial \omega_t}s(\omega_t) + \frac{1}{2} \frac{\partial^2 \chi_t}{\partial \W_t^2} \right. + \frac{\partial^2 \chi_t}{\partial \W_t \partial \omega_t}j(\omega_t)\\
&\!\!\!\!\left.+\frac{1}{2} j(\omega_t)^{\top}\frac{\partial^2 \chi_t}{\partial \omega_t^2} j(\omega_t) \right]dt + \left[\frac{\partial \chi_t}{\partial \W_t} + \frac{\partial \chi_t}{\partial \omega_t}j(\omega_t) \right]d\W_t,
\end{array}
\end{equation}
for all $\omega_t \in W(w)$. In addition, the steady-state response of system~(\ref{MRNS-eqInt}) is $x^{ss}_t=\chi_t(\W_t,\omega_t)$ almost surely for any $(\omega_0,x_0)\in W(w)\times X(w)$, where $X(w)\subset E_s(w)$ is a neighborhood of the origin.
\end{theorem}

\textit{Proof:}
Consider the interconnection~(\ref{MRNS-eqInt}) of system~(\ref{MRNS-eq1}) and the signal generator~(\ref{MRNS-eq2}). Computing the linearization at the zero equilibrium of this system yields
$$
\left[\!\! \begin{array}{c}d\bar{\omega}_t \\ d\bar{x}_t \end{array}\!\!\right]= \left[ \begin{array}{cc}S & 0\\ \tilde{B} & A \end{array}\right]\left[ \begin{array}{c}\bar{\omega}_t \\ \bar{x}_t\end{array}\right]dt +\left[ \begin{array}{cc}J & 0\\\tilde{G} & F \end{array}\right]\left[ \begin{array}{c}\bar{\omega}_t \\ \bar{x}_t \end{array}\right]d\W_t,
$$
where 
$$
\left.A=\frac{\partial f(x_t,l(\omega_t))}{\partial x_t}\right|_{\substack{x_t=0\\\omega_t=0}}, \quad \left.\tilde{B}=\frac{\partial f(x_t,l(\omega_t))}{\partial \omega_t}\right|_{\substack{x_t=0\\\omega_t=0}}, 
$$
$$
\left.F=\frac{\partial g(x_t,l(\omega_t))}{\partial x_t}\right|_{\substack{x_t=0\\\omega_t=0}}, \quad \left.\tilde{G}=\frac{\partial g(x_t,l(\omega_t))}{\partial \omega_t}\right|_{\substack{x_t=0\\\omega_t=0}}, 
$$ 
$$
\left.S=\frac{\partial s(\omega_t)}{\partial \omega_t}\right|_{\omega_t=0}, \quad \left.J=\frac{\partial j(\omega_t)}{\partial \omega_t}\right|_{\omega_t=0}.
$$
By Assumptions~\ref{MRNS-as1} and \ref{MRNS-as2}, Theorem~\ref{ORS-thLSSX} applies to this linearized system. Thus, by appealing to the center manifold theory (see \cite[Section 2.3]{Car:81} for deterministic systems, and \cite{Box:89} for a stochastic version), there exists a neighborhood $W(w)\times X(w)$ of $(0,0)$ in which the interconnected system has a center manifold at $(\omega_t,x_t)=(0,0)$ described by the graph of $x_t=\chi_t(\W_t,\omega_t)$. By replacing $\chi_t$ in (\ref{MRNS-eq1}) we obtain the characterization given by equation~(\ref{MRNS-eqNSS}), with the property that $\textstyle\left.\X_t=\frac{\partial \chi_t}{\partial \omega_t}\right|_{\omega_t=0}$, where $\X_t$ is the solution of (\ref{ORS-eqPI}). In addition, by \cite[Theorem 7.1(i)]{Box:89} the stochastic center manifold is locally exponentially attractive almost surely \textit{i.e.} for all pairs $(\omega_0^\star,x_0^{\circ})$ in the neighborhood $W(w)\times X(w)$ of $(0,0)$, the inequality\footnote{The definition of the norm $|\cdot|_w$ is given in the Appendix (Definition~\ref{def-rn}).}
$$
| x_t-\chi_t(\W_t,\omega_t)|_w \le K(t) | x_0^{\circ}-\chi_0(0,\omega_0^\star)|_w,
$$
with $K$ such that $\textstyle \lim_{t \to +\infty}\frac{1}{t}\log K(t)<0$, holds for all $t\ge 0$ almost surely. This together with the invariance of the stochastic center manifold proves that the steady state of system~(\ref{MRNS-eq1}) driven by~(\ref{MRNS-eq2}) is described almost surely by $x_t^{ss}=\chi_t(\W_t,\omega_t)$ for any $(\omega_0,x_0)\in W(w)\times X(w)$. Finally, equation~(\ref{MRNS-eqSPDE}) for the differential $d\chi_t$ is a direct consequence of It\^{o}'s formula (recalled as Lemma~\ref{MRSS-lmIto} in the Appendix).
\hspace*{0pt}\hfill $\square$ \par

\begin{remark}
If $g\equiv 0$ and $j\equiv 0$, then $\chi_t$ is a deterministic mapping. In this case, all the partial derivatives in (\ref{MRNS-eqSPDE}) are zero apart for $\frac{\partial \chi_t}{\partial \omega_t}s(\omega_t)$. Thus, by setting $\pi=\chi_t$, where $\pi:\omega\mapsto \pi(\omega)$, equation~(\ref{MRNS-eqNSS}) reduces to the well-known partial differential equation \cite[Equation~(8.3)]{Isi:95}
\begin{equation}
\label{MRSSeqISI}
\frac{\partial \pi}{\partial \omega}s(\omega) = f(\pi(\omega),l(\omega)).
\end{equation}
\end{remark}

\begin{remark}
Similarly to Theorem~\ref{ORS-thLSSX}, Theorem~\ref{MRNS-thm1} establishes that among the solutions of~(\ref{MRNS-eqNSS}), there exists at least one attractive solution that can be used to describe the steady state of the interconnection of system~(\ref{MRNS-eq1}) with the signal generator~(\ref{MRNS-eq2}).
\end{remark}

Inspired\footnote{The reason for using the word ``moment'' in the proposed stochastic generalization is that if we restrict the system to be linear and deterministic, \textit{i.e} if $f(x,u)=Ax+Bu$, $g(x,u) = 0$ and $h(x) =Cx$, then the generalized definition collapses into the classical definition of moment.} by \cite[Chapter 2.2]{ScaAst:17}, we can now define the moment in the stochastic framework.

\begin{definition}
\label{MRSS-DefMom}
Consider system~(\ref{MRNS-eq1}) and the signal generator~(\ref{MRNS-eq2}). The \textit{moment of system~(\ref{MRNS-eq1}) at $(s,j,l)$} is the mapping $h\circ \chi_t$, where $\chi_t$ is given in Theorem~\ref{MRNS-thm1}.
\end{definition}
\begin{remark}
The mapping $h\circ \chi_t$ is a function of time and of the stochastic process $\W_t$. Moments as functions of time have been introduced in \cite{ScaAst:15c} for discontinuous signal generators, in \cite{ScaAst:16d} for some classes of hybrid systems and in \cite{ScaTeeAst:17} for linear differential inclusions. Hence, the moment of system~(\ref{MRNS-eq1}), which sometimes we call ``stochastic moment'', generalizes both the classical time-invariant moments (see \textit{e.g.} \cite{Ast:10,ScaAst:15b}) and the time-varying moments introduced in those papers. 
\end{remark}

The differential defined in (\ref{MRNS-eqSPDE}) looks intimidating. For the sake of providing a worked example on how to manipulate (\ref{MRNS-eqSPDE}), we now show how to obtain~(\ref{ORS-eqPI}) directly from (\ref{MRNS-eqSPDE}) when the system and the generator are linear. 

\begin{example}
Consider system~(\ref{ORS-eq1}) and the generator~(\ref{ORS-eq2}) and assume $\chi_t=\X_t \omega_t$. In this example we show that equation~(\ref{MRNS-eqNSS}) implies equation~(\ref{ORS-eqPI}). First of all, equation~(\ref{MRNS-eqSPDE}) becomes
$$
\begin{array}{rl}
\!\!d\chi_t&\!\!\!\!= \left[\frac{\partial \X_t}{\partial t}\omega_t + \X_t S \omega_t + \frac{1}{2}\frac{\partial^2 \X_t}{\partial \W_t^2}\omega_t + \frac{\partial \X_t}{\partial \W_t}J\omega_t + 0\right]dt\\
&\!\!\!\!+\left[\frac{\partial \X_t}{\partial \W_t}\omega_t +\X_t J\omega_t \right]d\W_t.
\end{array}
$$
By It\^{o}'s Lemma the stochastic process $\X_t$ has the differential
$$
d\X_t = \left[\frac{\partial \X_t}{\partial t} + \frac{1}{2}\frac{\partial^2 \X_t}{\partial \W_t^2}\right]dt + \frac{\partial \X_t}{\partial \W_t}d\W_t.
$$
Comparing the previous two equations yields
$$
d\chi_t=d\X_t\omega_t + \left[\X_t S \omega_t + \frac{\partial \X_t}{\partial \W_t}J\omega_t\right]dt + \X_t J\omega_t d\W_t.
$$
Thus, equation (\ref{MRNS-eqNSS}) becomes
$$
\begin{array}{rl}
d\X_t\omega_t  \!\!\!\!&=\left(A\X_t-\X_t S+BL-\frac{\partial \X_t}{\partial \W_t}J\right)\omega_tdt\\[2mm]
&+\left(F\X_t-\X_t J+GL\right)\omega_td\W_t.
\end{array}
$$
By factoring out $\omega_t$ and noticing that the previous equation implies $\textstyle\frac{\partial \X_t}{\partial \W_t}=F\X_t-\X_t J+GL$, yields~(\ref{ORS-eqPI}).
\end{example}

Note that there are simpler ways to achieve the same result, as pointed out in the next remark.

\begin{remark}
The previous result can be obtained directly from $d(\X_t\omega_t)$ using the stochastic product rule. In fact, by Lemma~\ref{MRSS-lmdd} we have
$$
\begin{array}{rl}
\!\!\!d(\X_t\omega_t)&\!\!\!\!\!=\!d\X_t\omega_t+\X_td\omega_t+d\X_td\omega_t\\
&\!\!\!\!\!=\!d\X_t\omega_t+\X_t(S\omega_t dt + J \omega_t d\W_t) + [\X_t]_{\W} J \omega_t dt,
\end{array}
$$
where the notation $[\X_t]_{\W}$ indicates the component of $d\X_t$ which multiplies $d\W_t$. Hence, (\ref{MRNS-eqNSS}) becomes
$$
\begin{array}{l}
\!\!\!d\X_t+ [\X_t]_{\W} J dt \\[2mm]
\qquad =(A\X_t-\X_t S + BL)dt+(F\X_t -\X_t J + GL)d\W_t
\end{array}
$$
from which it follows that $[\X_t]_{\W}=F\X_t-\X_t J+GL$, thus proving the claim. 
\end{remark}

This last remark provides us with a quick analytical approach to reformulate the results of Theorem~\ref{ORS-thLSSX} in the case in which there are multiple uncorrelated Brownian motions. To illustrate this possibility, consider the system described by
\begin{equation}
\label{ORS-eq1u}
dx_t=[Ax_t+Bu_t]dt+[Fx_t+Gu_t]d\W_t^x
\end{equation}
and the generator
\begin{equation}
\label{ORS-eq2u}
d\omega_t= S\omega_t dt+J\omega_t d\W_t^s,
\end{equation}
where $\W_t^x$ and $\W_t^s$ are two uncorrelated Brownian motions, \textit{i.e.} $d\W_t^x d\W_t^s=0$.

\begin{corollary}
\label{ORS-coLSS}
Consider system~(\ref{ORS-eq1u}) driven by the signal generator~(\ref{ORS-eq2u}). The process $\X_t$ in Theorem~\ref{ORS-thLSSX} is now the steady-state solution of
\begin{equation}
\label{MRSS-eqLSSu}
\begin{array}{rl}
\!\!\!d\X_t\!=\!\!\!\!\!&\left(A\X_t-\X_t\left(S-J^2\right)+BL\right)dt+\left(F\X_t+GL\right)d\W_t^x\\[2mm]
& -\X_t J d\W_t^s.
\end{array}
\end{equation}
\end{corollary}

\textit{Proof:} We focus on the derivation of equation~(\ref{MRSS-eqLSSu}). For now, assume that the steady state of system~(\ref{ORS-eq1u}) driven by the signal generator~(\ref{ORS-eq2u}) can be written as $x_t^{ss}=\X_t \omega_t$ for some $\X_t$. We use the stochastic product rule to compute $d(\X_t\omega_t)$ yielding
$$
\begin{array}{rl}
d(\X_t\omega_t)&\!\!\!\!=d\X_t\omega_t+\X_td\omega_t+d\X_td\omega_t\\
&\!\!\!\!=d\X_t\omega_t+\X_t(S\omega_t dt + J \omega_t d\W_t^s) + d\X_td\omega_t.
\end{array}
$$
At steady state (\textit{i.e.} $x_t^{ss}=\X_t \omega_t$) this last equation, (\ref{ORS-eq1u}) and (\ref{ORS-eq2u}) give
$$
\begin{array}{rl}
\!\!\!d\X_t\omega_t+d\X_td\omega_t&\!\!\!\!=\left(A\X_t-\X_t S+BL\right)\omega_tdt\\
&+\left(F\X_t+GL\right)\omega_td\W_t^x-\X_t J \omega_td\W_t^s.
\end{array}
$$
The only non-zero term in $d\X_td\omega_t$ is the one resulting from the product of the components of $d\X_t$ and $d\omega_t$ which multiply $d\W_t^s$. Hence, we easily see that $d\X_td\omega_t=-\X_t J^2 \omega_t dt$, from which~(\ref{MRSS-eqLSSu}) follows. Repeating the steps of the proof of Theorem~\ref{ORS-thLSSX}, \textit{i.e.} defining $z_t:=x_t - \X_t\omega_t$ and computing $dz_t$, it is straightforward to show that the relation $x_t^{ss}=\X_t \omega_t$ indeed holds. 
\hspace*{0pt}\hfill $\square$ \par

Combining the results of Theorem~\ref{ORS-thLSSX} and Corollary~\ref{ORS-coLSS}, it is straightforward to generalize the theory to an arbitrary number of identical, correlated or uncorrelated Brownian motions.

\section{reduced-order models}
\label{MRNS-secROM}

In this section we provide families of linear and nonlinear reduced-order models solving Problem~\ref{MRSto-pro}. We present first the results for nonlinear systems and then for linear systems.

\subsection{Nonlinear systems}

In accordance with Problem~\ref{MRSto-pro}
we provide the definition of (reduced) model of system~(\ref{MRNS-eq1}) at $(s,j,l)$.
\begin{definition}
\label{MRNS-defROM}
Consider system~(\ref{MRNS-eq1}) and the signal generator~(\ref{MRNS-eq2}). The system described by equation~(\ref{MRSto-eqRM})
is a \textit{stochastic model of system~(\ref{MRNS-eq1}) at $(s,j,l)$} if system~(\ref{MRSto-eqRM}) has the same moment at $(s,j,l)$ of system~(\ref{MRNS-eq1}). System~(\ref{MRSto-eqRM}) is a \textit{stochastic reduced-order model of system~(\ref{MRNS-eq1}) at $(s,j,l)$} if $\nu<n$.
\end{definition}

From this definition a result follows straightforwardly.
\begin{lemma}
\label{MRNS-lmROM1}
Consider system~(\ref{MRNS-eq1}) and the signal generator~(\ref{MRNS-eq2}). Suppose Assumptions~\ref{MRNS-as1} and \ref{MRNS-as2} hold. Then system~(\ref{MRSto-eqRM}) is a stochastic model of system~(\ref{MRNS-eq1}) at $(s,j,l)$ if there exists a stochastic process $\rho:(t,\W_t,\omega_t)\mapsto \rho_t(\W_t,\omega_t)$, locally defined in a neighborhood $\red{W}(w) \subset E_c(w)$ of $\omega_t=0$, with $\rho_t(0,0)=0$, which satisfies the equation
\begin{equation}
\label{MRNS-eqP}
\begin{array}{l}
d \rho_t= \red{f}(\rho_t(\W_t,\omega_t),l(\omega_t))dt + \red{g}(\rho_t(\W_t,\omega_t),l(\omega_t))d\W_t,
\end{array}
\end{equation}
and it is such that for all $t\ge0$
\begin{equation}
\label{MRNS-eqCP}
h(\chi_t(\W_t,\omega_t))=\red{h}(\rho_t(\W_t,\omega_t)),
\end{equation}
almost surely, where $\chi_t$ is a solution of (\ref{MRNS-eqNSS}).
\end{lemma}

\textit{Proof:}
If $\rho_t$ is a solution of equation~(\ref{MRNS-eqP}), then the moment of system~(\ref{MRSto-eqRM}) at $(s,j,l)$ is by definition the mapping $\red{h} \circ \rho_t$. Equation~(\ref{MRNS-eqCP}) imposes that the moment of system~(\ref{MRSto-eqRM}) is equal to the moment of system~(\ref{MRNS-eq1}). \hspace*{0pt}\hfill $\square$ \par

Problem~\ref{MRSto-pro} can be now reformulated as the problem of determining the mappings $\red{f}$, $\red{g}$ and $\red{h}$ in Lemma~\ref{MRNS-lmROM1} such that the two equations~(\ref{MRNS-eqP}) and (\ref{MRNS-eqCP}) are satisfied. This problem can be solved easily. 


\begin{proposition}
\label{MRNS-lmROM2}
Consider system~(\ref{MRNS-eq1}) and the signal generator~(\ref{MRNS-eq2}). Suppose Assumptions~\ref{MRNS-as1} and \ref{MRNS-as2} hold. Then the system
\begin{equation}
\label{MRNS-eqROM2}
\begin{array}{rl}
d \red{x}_t &\!\!\!\!=(s(\red{x}_t) -\delta(\red{x}_t)l(\red{x}_t) + \delta(\red{x}_t) u_t)dt \\[2mm]
&+(j(\red{x}_t)-\eta(\red{x}_t)l(\red{x}_t)+\eta(\red{x}_t)u_t)d\W_t, \\[2mm]
\red{y}&\!\!\!\!=h(\chi_t(\W_t,\red{x}_t)),
\end{array}
\end{equation}
where $\chi_t$ is a solution of (\ref{MRNS-eqNSS}), is a stochastic model of system~(\ref{MRNS-eq1}) at $(s,j,l)$ if $\delta$ and $\eta$ are arbitrary mappings such that equation
\begin{equation}
\label{MRNS-eqrho111}
\begin{array}{l}
d\rho_t = \left(s(\rho_t) - \delta(\rho_t)l(\rho_t) +\delta(\rho_t)l(\omega_t)\right) dt \\[2mm]
\qquad +
\left(j(\rho_t) -\eta(\rho_t)l(\rho_t)+\eta(\rho_t)l(\omega_t)\right)d\W_t,
\end{array}
\end{equation}
has the trivial solution $\rho_t(\W_t,\omega_t)=\omega_t$.
\end{proposition}

\textit{Proof:}
In model~(\ref{MRSto-eqRM}) we select $\red{f}$ and $\red{g}$ such that the system is affine in the input, namely $\red{f}(\red{x}_t,u_t)=\phi(\red{x}_t)+\delta(\red{x}_t) u_t$ and $\red{g}(\red{x}_t,u_t)=\gamma(\red{x}_t)+\eta(\red{x}_t) u_t$ for some smooth mapping $\phi$, $\delta$, $\gamma$ and $\eta$ to be determined. Selecting $\rho_t(\W_t,\omega_t)=\omega_t$, \textit{i.e.} the identity mapping, yields (by writing (\ref{MRNS-eqSPDE}) for the reduced-order model) $d\rho_t = s(\omega_t) dt + j(\omega_t)d\W_t$. Hence, equation~(\ref{MRNS-eqP}) becomes
$$
\begin{array}{rl}
s(\omega_t) dt + j(\omega_t)d\W_t&\!\!\!\!= \left( \phi(\omega_t)+\delta(\omega_t) l(\omega_t)\right) dt \\
&+ \left( \gamma(\omega_t)+\eta(\omega_t) l(\omega_t)\right) d\W_t
\end{array}
$$
which is satisfied selecting
$$
\phi(\omega_t)=s(\omega_t)-\delta(\omega_t) l(\omega_t), \qquad
\gamma(\omega_t)=j(\omega_t)-\eta(\omega_t) l(\omega_t),
$$
\textit{i.e.} (\ref{MRNS-eqrho111}) is satisfied.
Thus, selecting $\phi(\red{x}_t) = \left. \phi(\omega_t)\right|_{\omega_t=\rho^{-1}(\red{x}_t)}$ and $\gamma(\red{x}_t) = \left. \gamma(\omega_t)\right|_{\omega_t=\rho^{-1}(\red{x}_t)}$,
model~(\ref{MRNS-eqROM2}) satisfies equations~(\ref{MRNS-eqP}) and (\ref{MRNS-eqCP}) for any mapping $\delta$ and $\eta$ such that~(\ref{MRNS-eqrho111})  has the unique solution $\rho_t(\W_t,\omega_t)=\omega_t$.
\hspace*{0pt}\hfill $\square$ \par

The family of models (\ref{MRNS-eqROM2}) is parametrized by the mappings $\delta$ and $\eta$. These mappings can be used to span the family of reduced-order models by moment matching so that specific additional properties are imposed. For instance, we may want to preserve stability or we may want to achieve a
special representation of the reduced-order model which is particularly
useful for a desired application. 

Note that in Definition~\ref{MRSS-DefMom} the moment of system~(\ref{MRNS-eq1}) is defined as the solution of an equation, namely~(\ref{MRNS-eqNSS}), without any reference to the steady state of the system. In fact, the moment may exist (when equation~(\ref{MRNS-eqNSS}) has a solution) even though Assumption~\ref{MRNS-as1} is not satisfied and we cannot identify the mapping $\chi_t$ as a steady state. In fact, Assumptions~\ref{MRNS-as1} and \ref{MRNS-as2} are sufficient to guarantee that equation~(\ref{MRNS-eqNSS}) has a solution, but not necessary. This fact explains why asymptotic stability of the origin of the reduced-order model is not required by Lemma~\ref{MRNS-lmROM1} and Proposition~\ref{MRNS-lmROM2}. However, one can use the mappings $\delta$ and $\eta$ to impose stability and simplify the result of Proposition~\ref{MRNS-lmROM2}.

\begin{corollary}
\label{MRNS-cor1}
Consider system~(\ref{MRNS-eq1}) and the signal generator~(\ref{MRNS-eq2}). Suppose Assumptions~\ref{MRNS-as1} and \ref{MRNS-as2} hold. Then system~(\ref{MRNS-eqROM2}) is a stochastic model of system~(\ref{MRNS-eq1}) at $(s,j,l)$ for any $\delta$ and $\eta$ such that the linearization of (\ref{MRNS-eqROM2}) around zero has all negative Lyapunov exponents almost surely.
\end{corollary}

\textit{Proof:}
This is a direct consequence of Proposition~\ref{MRNS-lmROM2} and Theorem~\ref{MRNS-thm1}.
\hspace*{0pt}\hfill $\square$ \par

Note that there exist multiple $\delta$ and $\eta$ satisfying the condition in Corollary~\ref{MRNS-cor1}. Thus, $\delta$ and $\eta$ are still partially free to be used to impose additional properties (\textit{e.g.} different speeds of decay of the transient among all stabilizing mappings $\delta$ and $\eta$).

\subsection{Linear systems}

This section has the same structure as the previous section. We begin by providing the definition of model of system~(\ref{ORS-eq1}) at $(S,J,L)$.
\begin{definition}
\label{MRLS-defROM}
Consider system~(\ref{ORS-eq1}) and the signal generator~(\ref{ORS-eq2}). The system described by the equations
\begin{equation}
\label{MRLS-eqROM}
d \red{x}_t =(\red{A} \red{x}_t + \red{B} u_t)dt + (\red{F} \red{x}_t +\red{G} u_t)d\W_t, \quad \red{y}_t=\red{C} \red{x}_t,
\end{equation}
where $\red{A}\in \mathbb{R}^{\nu\times \nu}$, $\red{B}\in \mathbb{R}^{\nu\times1}$, $\red{F}\in \mathbb{R}^{\nu\times \nu}$, $\red{G}\in \mathbb{R}^{\nu\times1}$, $\red{C}\in \mathbb{R}^{1\times \nu}$, is a \textit{stochastic model of system~(\ref{ORS-eq1}) at $(S,J,L)$}, if system~(\ref{MRLS-eqROM}) has the same moments of system~(\ref{ORS-eq1}) at $(S,J,L)$. System~(\ref{MRLS-eqROM}) is a \textit{stochastic reduced-order model of system~(\ref{ORS-eq1}) at $(S,J,L)$} if $\nu<n$.
\end{definition}

The equivalent of Lemma~\ref{MRNS-lmROM1} follows straightforwardly.
\begin{lemma}
\label{MRLS-lmROM1}
Consider system~(\ref{ORS-eq1}) and the signal generator~(\ref{ORS-eq2}). Suppose Assumptions~\ref{ORS-asSys} and \ref{ORS-asGen} hold. Then system~(\ref{MRLS-eqROM}) is a stochastic model of system~(\ref{ORS-eq1}) at $(S,J,L)$ if there exists $\cR_t\in\mathbb{R}^{\nu \times \nu}$
which satisfies the equation
\begin{equation}
\label{MRLS-eqP}
\begin{array}{rl}
d\cR_t=\!\!\!\!&\left(\red{A}\cR_t-\cR_t\left(S-J^2\right)-\red{F}\cR_t J+\red{B}L - \red{G}LJ \right)dt\\[2mm]
&+(\red{F}\cR_t-\cR_tJ+\red{G}L)d\W_t,
\end{array}
\end{equation}
and it is such that for all $t\ge 0$
\begin{equation}
\label{MRLS-eqCP}
C\X_t=\red{C} \cR_t,
\end{equation}
almost surely, where $\X_t$ is a solution of (\ref{ORS-eqPI}).
\end{lemma}

\textit{Proof:} The proof is analogous to the proof of Lemma~\ref{MRNS-lmROM1} and thus is omitted.
\hspace*{0pt}\hfill $\square$ \par

In the spirit of Proposition~\ref{MRNS-lmROM2} we select the matrices of the model to satisfy equations~(\ref{MRLS-eqP}) and (\ref{MRLS-eqCP}). 

\begin{proposition}
\label{MRLS-lmROM2}
Consider system~(\ref{ORS-eq1}) and the signal generator~(\ref{ORS-eq2}). Suppose Assumptions~\ref{ORS-asSys} and \ref{ORS-asGen} hold. Then the system
\begin{equation}
\label{MRLS-eqROM2}
\begin{array}{rl}
d \red{x}_t &\!\!\!\!=((S-\red{B}L) \red{x}_t + \red{B} u_t)dt + ((J-\red{G} L) \red{x}_t + \red{G} u_t)d\W_t, \\[2mm]
\red{y}_t&\!\!\!\!=C\X_t \red{x}_t,
\end{array}
\end{equation}
where $\X_t$ is a solution of (\ref{ORS-eqPI}), is a stochastic model of system~(\ref{ORS-eq1}) at $(S,J,L)$ for any $\red{B}$ and $\red{G}$ such that equation~(\ref{MRLS-eqP}) has the unique solution $\cR_t = I$.
\end{proposition}

\textit{Proof:} Equation~(\ref{MRLS-eqP}) has solution $\cR_t = I$ only if
$$
\begin{array}{l}
0 = \red{A}-\left(S-J^2\right)-\red{F}J+\red{B}L - \red{G}LJ,\\
0 =\red{F}-J+\red{G}L.
\end{array}
$$
Solving the second equation with respect to $\red{F}$ yields $\red{F}=J-\red{G}L$. Replacing this expression in the first equation and solving with respect to $\red{A}$ yields $\red{A}=S-\red{B}L$. Finally, substituting $\cR_t = I$ in equation~(\ref{MRLS-eqCP}) yields $\red{C} =C\X_t$.
\hspace*{0pt}\hfill $\square$ \par

Note that the matrices $\red{B}$ and $\red{G}$ play the same role that $\delta$ and $\eta$ have in the nonlinear case. Thus $\red{B}$ and $\red{G}$ can be freely selected to achieve additional properties for the reduced-order model.

Note that there is no loss of generality in selecting the solution of~(\ref{MRLS-eqP}) as in Proposition~\ref{MRLS-lmROM2}. In fact, if a model~(\ref{MRLS-eqROM}) satisfies Lemma~\ref{MRLS-lmROM1} for $\cR_t \ne I$, we can define a stochastic change of coordinates for which the model in the new coordinates has the form~(\ref{MRLS-eqROM2}). We formalize this property in the next result.
\begin{lemma}
Assume that system~(\ref{MRLS-eqROM}) is a reduced-order model satisfying Lemma~\ref{MRLS-lmROM1}. Let $\xi_t:= \cR_t^{-1}\red{x}_t$. Then the system
\begin{equation}
\label{MRLS-eqROM2bis}
\begin{array}{rl}
d \xi_t &\!\!\!\!=((S-\overline{B}L) \xi_t + \overline{B} u_t)dt + ((J-\overline{G} L) \xi_t + \overline{G} u_t)d\W_t, \\[2mm]
\red{y}_t&\!\!\!\!=C\X_t \xi_t,
\end{array}
\end{equation}
with $\overline{G}=\cR_t^{-1}\red{G}$ and $\overline{B}=\cR_t^{-1}(\red{B}-(\red{F}\cR_t-\cR_tJ+\red{G}L)\overline{G})$ satisfies Proposition~\ref{MRLS-lmROM2}.
\end{lemma}

\textit{Proof:} Note that $d\red{x}_t=d\cR_t \xi_t +\cR_t d\xi_t+d\cR_t d\xi_t$ and $d\xi_t=\cR_t^{-1}\left((\red{A} \red{x}_t + \red{B} u_t)dt + (\red{F} \red{x}_t +\red{G} u_t)d\W_t-d\cR_t (\xi_t+ d\xi_t)\right)$. The result follows from tedious but straightforward computation.
\hspace*{0pt}\hfill $\square$ \par

\section{Classes of approximated reduced-order models}
\label{MRSS-sec-AROM}

Analysing the models proposed in Propositions~\ref{MRNS-lmROM2} and \ref{MRLS-lmROM2}, we note that the method relies on the determination of the mappings $\chi_t$ and $\X_t$, respectively. Currently, the determination of these mappings requires the solution of equations~(\ref{ORS-eqPI}) and (\ref{MRNS-eqNSS}), respectively, which is computationally expensive. For instance, equation~(\ref{ORS-eqPI}) is a stochastic matrix equation which consists of $n\nu$ linear stochastic equations. Thus, to determine $\red{y}_t$ we need to construct $n\nu$ stochastic processes, \textit{i.e.} the components of $\X_t$. This would not be a problem if $\X_t$ could be determined off-line, but since $\X_t$ depends on the Brownian motion, this is not possible. Hence, while the models proposed in Proposition~\ref{MRLS-lmROM2} possess the same moment or, equivalently, the same steady-state output response of system~(\ref{ORS-eq1}), they can be hardly considered ``simpler''. Note that this issue could eventually be solved. In fact, on one hand what we need is $C\X_t$, which is of order $\nu$, rather than $\X_t$. Thus, the development of efficient ways of determining the moment $C\X_t$ without computing $\X_t$ may solve the issue (this is normally achieved in the deterministic case, see \textit{e.g.} \cite{Ant:05}). On the other hand, it may even be possible to develop methods to determine directly models that match the moment without computing it (as achieved by some deterministic techniques, see \textit{e.g.} \cite{GugAntBea:08}).

In this paper we propose an alternative way to overcome this difficulty. In particular, the idea is to determine stochastic reduced-order models that, maintaining a subset of the stochastic properties of the system to be reduced, allows us to carry out off-line all computations that have a complexity which depends on $n$, thus providing a computational advantage. In the following we propose classes of ``approximated'' reduced-order models and we discuss the properties of these models. For simplicity, the rest of the paper focuses on linear systems, although nonlinear extensions are briefly mentioned.

\subsection{Reduced-order models preserving the mean of the moment $E[C\X_t]$}

The first class of models that we consider are models obtained considering a relaxation of Problem~\ref{MRSto-pro} in which the limit~(\ref{MRSto-eqMMc}), namely
$$
\lim_{t\to \infty} C\X_t\omega_t - \red{C}\cR_t\omega_t = 0
$$
is replaced by
\begin{equation}
\label{MRLS-eqCe2}
\lim_{t\to \infty} CE[\X_t]\omega_t - \red{C}E[\cR_t]\omega_t = 0.
\end{equation}
In other words, we approximate $\X_t$ with its expectation and, instead of matching the moment, we match just its expectation. To make the concept precise, we introduce the following definition.

\begin{definition}
\label{MRLS-defROMm}
Consider system~(\ref{ORS-eq1}) and the signal generator~(\ref{ORS-eq2}). The system described by equation~(\ref{MRLS-eqROM}) is a \textit{stochastic model in the moment-mean of system~(\ref{ORS-eq1}) at $(S,J,L)$}, if the mean of the moment of system~(\ref{MRLS-eqROM}) is equal to the mean of the moment of system~(\ref{ORS-eq1}) at $(S,J,L)$. System~(\ref{MRLS-eqROM}) is a \textit{stochastic reduced-order model in the moment-mean of system~(\ref{ORS-eq1}) at $(S,J,L)$} if $\nu<n$.
\end{definition}

To determine a class of models that satisfy Definition~\ref{MRLS-defROMm}, let $\Pi= E[\X_t]$ and note, from~(\ref{ORS-eqPI}), that $\Pi$ obeys the equation
\begin{equation}
\label{MRSS-eqEXi}
d\Pi= \left(A\Pi-\Pi\left(S-J^2\right)-F\Pi J+BL-GLJ\right)dt.
\end{equation}
The equilibrium point of this equation is given by solving the generalized Sylvester equation\footnote{This equation has a unique solution if and only if $0 \not \in \sigma\left(I\otimes A - (S-J^2)^{\top} \otimes I - J^{\top} \otimes F \right)$.}
\begin{equation}
\label{MRSS-eqEPI}
A\Pi-\Pi\left(S-J^2\right)-F\Pi J+BL-GLJ = 0.
\end{equation}
The equilibrium point is unique and describes the steady-state solution of~(\ref{MRSS-eqEXi}) if $A$, $F$, $S$ and $J$ are such that the equilibrium of (\ref{MRSS-eqEXi}) is asymptotically stable.\\
A family of reduced-order models in the moment-mean follows.

\begin{proposition}
\label{MRLS-lmROM3}
Consider system~(\ref{ORS-eq1}) and the signal generator~(\ref{ORS-eq2}). Suppose Assumptions~\ref{ORS-asSys} and \ref{ORS-asGen} hold and that equation~(\ref{MRSS-eqEXi}) has an asymptotically stable equilibrium point $\Pi$ solving equation~(\ref{MRSS-eqEPI}). Then the system
\begin{equation}
\label{MRLS-eqROM3}
\begin{array}{rl}
d \red{x}_t &\!\!\!\!=((S-\red{B}L) \red{x}_t + \red{B} u_t)dt + (\red{F} \red{x}_t + \red{G} u_t)d\W_t, \\[2mm]
\red{y}_t&\!\!\!\!=C\Pi \red{x}_t,
\end{array}
\end{equation}
is a stochastic model in the moment-mean of system~(\ref{ORS-eq1}) at $(S,J,L)$, for any $\red{G}$ and for any $\red{B}$ and $\red{F}$ such that 
\begin{equation}
\label{MRLS-eqCER}
\sigma\left(I \otimes \left(S-\red{B}L\right) - \left(S-J^2\right)^{\top} \otimes I - J^{\top} \otimes \red{F} \right) \subset \mathbb{C}_{<0}.
\end{equation}
\end{proposition}

\textit{Proof:} Consider system~(\ref{MRLS-eqROM}) and the equation~(\ref{MRLS-eqP}) defining $\cR_t$. Let $R:=E[\cR_t]$ and note that $R$ satisfies
\begin{equation}
\label{MRLS-eqER}
dR=\left(\red{A}R-R\left(S-J^2\right)-\red{F}R J+\red{B}L - \red{G}LJ \right)dt.
\end{equation}
Let $\red{A}=S-\red{B}L$. If $\red{B}$ and $\red{F}$ are such that (\ref{MRLS-eqCER}) holds, then $R=I$ is the unique attractive equilibrium of equation~(\ref{MRLS-eqER}). Finally, the selection $\red{C}=C\Pi$ satisfies condition~(\ref{MRLS-eqCe2}).
\hspace*{0pt}\hfill $\square$ \par

\begin{remark}
The advantage of reduced-order models in the moment-mean is that we need to determine only $C\Pi$, which can be computed with a plethora of efficient methods (see \cite{Ant:05} for a review and \textit{e.g.} \cite{ScaAst:16a}).
\end{remark}

In the next simple example we illustrate the differences between a reduced-order model~(\ref{MRLS-eqROM2}) and a reduced-order model in the moment-mean~(\ref{MRLS-eqROM3}).

\begin{example}
\begin{figure}
\centering
\includegraphics[width=\columnwidth,height=6cm]{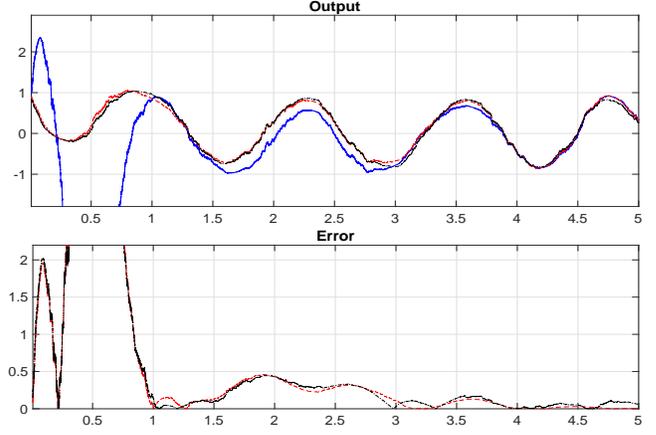}
\caption{Top graph: time history of the output of system~(\ref{ORS-eq1}) (solid/blue), of the output of the stochastic reduced-order model~(\ref{MRLS-eqROM2}) (dashed/red) and of the output of the stochastic reduced-order model in the moment-mean~(\ref{MRLS-eqROM3}) (dash-dotted/black). Bottom graph: time history of the corresponding absolute errors.}
\label{MRSS-fig1}
\end{figure}%
\begin{figure}
\centering
\includegraphics[width=\columnwidth]{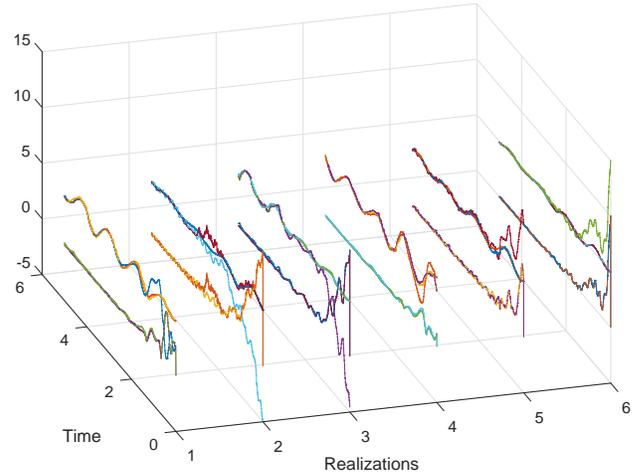}
\caption{Six different realizations of the simulation in Figure~\ref{MRSS-fig1}. For each realization, the top set of curves are the output trajectories and the bottom set are the corresponding absolute errors.}
\label{MRSS-fig1bis}
\end{figure}%
Consider a linear system~(\ref{ORS-eq1}) of order $n=200$ with randomly generated matrices\footnote{All the matrices used in this simulation can be downloaded from \cite{FILE_MRSS}.} $A$, $B$ and $C$, and with $F=0.05A$ and $G=0.1B$. Consider a signal generator~(\ref{ORS-eq2}) of order $\nu=2$ with $J$ a randomly generated such that $S=[0\,5;-5\,0]+0.5J^2$ and $J$ commute (thus Assumption~\ref{ORS-asGen} is verified by construction). A stochastic reduced-order model~(\ref{MRLS-eqROM2}) and a stochastic reduced-order model in the moment-mean~(\ref{MRLS-eqROM3}) are computed. For both models we have selected $\red{B}$ and $\red{G}$ such that the eigenvalues of $\red{A}=S-\red{B}L$ and $\red{F}=J-\red{G}L$ are two of the eigenvalues of $A$ and $F$, respectively. The top graph in Fig.~\ref{MRSS-fig1} shows the output of the system (solid/blue), of the stochastic reduced-order model~(\ref{MRLS-eqROM2}) (dashed/red) and of the stochastic reduced-order model in the moment-mean~(\ref{MRLS-eqROM3}) (dash-dotted/black). The bottom graph shows the corresponding absolute errors. We notice that both reduced-order models approximate the output of the system, although the model in the moment-mean shows a steady-state mismatch 
Fig.~\ref{MRSS-fig1bis} shows the same quantities for six different realizations. The top set of lines correspond to the top graph in Fig.~\ref{MRSS-fig1}, whereas the bottom set of lines correspond to the bottom graph in Fig.~\ref{MRSS-fig1}. The figure shows self-consistence across multiple realizations of the noise.
\end{example}

\subsection{Reduced-order models preserving the mean and mean-square of the steady state when $J=0$}

In this section we clarify the relation between the stochastic models in the moment-mean defined in the previous section and families of approximated stochastic models introduced in \cite{ScaTee:17,ScaTee:17a}. Subsequently, we provide additional results on the latter.

\subsubsection{Relation between models in the moment-mean and models in the mean introduced in \cite{ScaTee:17,ScaTee:17a}}

Since $\omega_t$ is a stochastic process that depends on the same noise of $\X_t$, we have that $E[\X_t\omega_t] \ne E[\X_t]E[\omega_t] \ne E[\X_t]\omega_t$. Hence, the reduced-order models in the moment-mean do not necessarily have the same steady-state mean of the output of the system to be reduced. Note, however, that if $J=0$, then $\omega_t$ is deterministic and condition~(\ref{MRLS-eqCe2}) can be written as
\begin{equation}
\label{MRLS-eqLEe}
\lim_{t\to \infty} E[e_t] = 0.
\end{equation}
Thus, in this case matching the mean of the moment corresponds to matching the mean of the steady-state output. Note also that the mean $E[C\X_t]=C\Pi$, where $\Pi$ is the unique solution of 
\begin{equation}
\label{MRLS-eqSyl}
A\Pi-\Pi S + BL = 0,
\end{equation}
is the moment of the deterministic system
$$
\dot m=Am+Bu,\qquad y=Cm,
$$
where $m=E[x_t]$. Thus, the stochastic models in the moment-mean defined in the previous section are a generalisation of the ``stochastic models in the mean'' introduced in \cite{ScaTee:17}, which were based on the idea of preserving the mean of the steady-state output of system~(\ref{ORS-eq1}) driven by~(\ref{ORS-eq2}), \textit{i.e.} satisfying~(\ref{MRLS-eqLEe}).

Note that, even though the moment $C\Pi$ is the moment of a deterministic system, the models in the moment-mean (and the models in the mean defined in \cite{ScaTee:17}) are in general stochastic models. 

\begin{remark}
\label{MRSS-rm7}
Consider the special selection $\red{F}=-\red{G}L$ in (\ref{MRLS-eqROM3}). The steady state of the resulting model for $u=L\omega$ is generated by a deterministic system. In fact, the equation
$$
d\cR_t=((S-\red{B}L)\cR_t-\cR_t S+\red{B}L)dt+(-\red{G}L\cR_t+\red{G}L)d\W_t,
$$
which is solved by $\cR_t=I$, is independent of $d\W_t$. Thus, at steady state we have that $-\red{G}L \red{x}^{ss}_t +\red{G}L \omega_t = -\red{G}L \cR_t \omega_t +\red{G}L \omega_t= 0$. As a result, the steady-state system is not stochastic. We stress, however, that this is not the case if $F_r\ne -G_rL$.
\end{remark}

Similarly, stochastic models in the mean can be defined also for nonlinear systems as those stochastic models that have the same mean of the steady-state output of system~(\ref{MRNS-eq1}) driven by~(\ref{MRNS-eq2}). A family of nonlinear stochastic models in the mean is given in the next result.

\begin{proposition}
\label{MRNS-lmROMNLMean}
Consider system~(\ref{MRNS-eq1}) and the signal generator~(\ref{MRNS-eq2}) with $j(\cdot)\equiv 0$. Suppose Assumptions~\ref{MRNS-as1} and \ref{MRNS-as2} hold and that $\sigma\left(\left.\frac{\partial f}{\partial x_t}\right|_{\substack{x_t=0\\\omega=0}}\right) \subset \mathbb{C}_{<0}$. Then the system
\begin{equation}
\label{MRNS-eqROM2Mean}
\begin{array}{rl}
d \red{x}_t &\!\!\!\!=(s(\red{x}_t) -\delta(\red{x}_t)l(\red{x}_t) + \delta(\red{x}_t) u)dt +\gamma(\red{x}_t,u)d\W_t, \\[2mm]
\red{y}&\!\!\!\!=h(\pi(\red{x}_t)),
\end{array}
\end{equation}
where $\pi$ is a solution of (\ref{MRSSeqISI}), is a stochastic model in the mean of system~(\ref{MRNS-eq1}) at $(s,l)$ for any $\gamma$ and for any $\delta$ such that the zero equilibrium of $d \red{x} =s(\red{x}) -\delta(\red{x})l(\red{x})$ is locally exponentially stable.
\end{proposition}

\textit{Proof:} The mean of system~(\ref{MRNS-eq1}) is described by $\dot{\overbrace{E[x_t]}} = f(E[x_t],u)$. This is a nonlinear deterministic system which has moment $h\circ \pi$, where $\pi$ is the solution of the deterministic partial differential equation~(\ref{MRSSeqISI}). Note that such solution exists by Assumption~\ref{MRNS-as2} and the assumption on the Jacobian of $f$, see \cite{Isi:95}. Then the claim follows from standard deterministic model reduction \cite{Ast:10}. 
\hspace*{0pt}\hfill $\square$ \par

\subsubsection{Stochastic models in the mean-square}
Although the models in the mean are stochastic systems, the matrices $\red{F}$ and $\red{G}$ in~(\ref{MRLS-eqROM3}) are free parameters and do not preserve, in a systematic way, information regarding the matrices $F$ and $G$ of the system to be reduced. Now, keeping the standing assumption that $J=0$, we want to ``improve'' the models in the mean. In particular, we want to use the parameters $\red{F}$ and $\red{G}$ to preserve the mean-square, in addition to the expectation, of the steady-state output of system~(\ref{ORS-eq1}) driven by~(\ref{ORS-eq2}). To preserve also this information we propose another class of reduced-order models that solve a variation of Problem~\ref{MRSto-pro} in which the limit~(\ref{MRSto-eqMMc}) is replaced by
\begin{equation}
\label{MRSS-Verr}
\lim_{t\to \infty} E[e_t] = 0 \quad \text{and} \quad \lim_{t\to \infty} E[e_te_t^{\top}] = 0.
\end{equation}
To make the concept precise, we introduce the following definition.
\begin{definition}
\label{MRLS-defROMV}
Consider system~(\ref{ORS-eq1}) and the signal generator~(\ref{ORS-eq2}). The system described by equation~(\ref{MRLS-eqROM}) is a \textit{stochastic model in the mean-square of system~(\ref{ORS-eq1}) at $(L,S)$}, if the conditions in (\ref{MRSS-Verr}) are simultaneously satisfied. System~(\ref{MRLS-eqROM}) is a \textit{stochastic reduced-order model in the mean-square of system~(\ref{ORS-eq1}) at $(L,S)$} if $\nu<n$.
\end{definition}

We remark that we have defined the models in the ``mean-square'' in such a way that in addition to preserving the mean-square of the steady state of system~(\ref{ORS-eq1}) driven by~(\ref{ORS-eq2}), these models preserve also its mean. To the end of determining reduced-order models in the mean-square we need a preliminary result, namely a description of the steady state of $M=E[x_t x_t^{\top}]$, which obeys the equation \cite[Theorem 4.5]{Gar:88}
\begin{equation}
\label{MRLS-eqVar}
\begin{array}{rl}
\dot M&=AM+MA^{\top}+FMF^{\top}+Bum^{\top}+m(Bu)^{\top}\\[2mm]
&\,\,\,\,\,\,+Fm(Gu)^{\top}+Gu(Fm)^{\top}+Gu(Gu)^{\top}.
\end{array}
\end{equation}

\begin{lemma}
\label{MRLS-lmSM}
Consider the interconnection of system~(\ref{MRLS-eqVar}) and the signal generator~(\ref{ORS-eq2}). Let $\mathcal{A}=I\otimes A + A\otimes I + F \otimes F$ and assume $\textstyle\sigma\left(\mathcal{A}\right)\subset\mathbb{C}_{<0}$ and $\sigma(S)\subset \mathbb{C}_{0}$. Then the steady-state response of such interconnection is
$$
\vect{M^{ss}}=\mathcal{K} \vect{\omega\omega^{\top}},
$$
where $\mathcal{K}$ is the unique solution of the augmented Sylvester equation
\begin{equation}
\label{MRLS-eqMSS}
\mathcal{A} \mathcal{K} + \mathcal{B}= \mathcal{K} \mathcal{S},
\end{equation}
with $\mathcal{S}=I\otimes S + S\otimes I$ and $\mathcal{B}=BL \otimes \Pi+ \Pi \otimes BL + GL \otimes F\Pi + F\Pi \otimes GL + GL \otimes GL$.
\end{lemma}

\textit{Proof.}
First of all note that $\mathcal{K}$ is the unique solution of the Sylvester equation~(\ref{MRLS-eqMSS}) because $\sigma(\mathcal{A})\cap\sigma(\mathcal{S})=\emptyset$. Let $\mu:=\vect{M}-\mathcal{K}\vect{\omega\omega^{\top}}$ and compute the derivative of $\mu$ with respect to time. Moreover, substitute $m$ with $\Pi\omega$, which is the steady state of $m$. Using the vectorization operator and the Kronecker product yields
$$
\dot \mu= \mathcal{A}\mu + (\mathcal{A} \mathcal{K} - \mathcal{K} \mathcal{S} + \mathcal{B})\vect{\omega\omega^{\top}}=\mathcal{A}\mu.
$$
Since $\textstyle\sigma\left(\mathcal{A}\right)\subset\mathbb{C}_{<0}$, $\mu$ converges exponentially to zero and $\vect{M}$ converges exponentially to $\mathcal{K}\vect{\omega\omega^{\top}}$.
\hspace*{0pt}\hfill $\square$ \par

We are now ready to give a family of reduced-order models in the mean-square.

\begin{proposition}
\label{MRLS-lmROM4}
Consider the interconnection of system~(\ref{ORS-eq1}) and the signal generator~(\ref{ORS-eq2}). Assume $\textstyle\sigma\left(\mathcal{A}\right)\subset\mathbb{C}_{<0}$ and $\sigma(S)\subset \mathbb{C}_{0}$. Assume there exists a matrix $\red{C}$ such that 
\begin{equation}
\label{MRSS-eqC1M}
\red{C} \otimes \red{C} = \left( C \otimes C \right) \mathcal{K}, 
\end{equation}
where $\mathcal{K}$ is the unique solution of~(\ref{MRLS-eqMSS}). Let $R$ be any invertible matrix such that $\red{C}R=C\Pi$, where $\Pi$ is the unique solution of~(\ref{MRLS-eqSyl}). Let
$\red{A}=RSR^{-1}-\red{B}LR^{-1}$ for any $\red{B}$ such that $\sigma(\red{A})\subset \mathbb{C}_{<0}$. Assume there exist matrices $\red{F}$ and $\red{G}$ such that
\begin{equation}
\label{MRLS-eqFrGr}
\begin{array}{l}
\!\red{F}\! \otimes\! \red{F}+\red{G}L \!\otimes\! \red{F} R + \red{F} R\! \otimes\! \red{G}L + \red{G} L \!\otimes\! \red{G} L=\\[2mm]
\,\,\,\,\,\,=-I\!\otimes\! \red{A} - \red{A}\!\otimes\! I+I\!\otimes \!S + S\!\otimes\! I - \red{B}L \!\otimes\! R- R\! \otimes\! \red{B}L,
\end{array}
\end{equation}
and that $\textstyle\sigma\left(I\otimes \red{A} + \red{A}\otimes I + \red{F} \otimes \red{F}\right)\subset\mathbb{C}_{<0}$. Then the system
\begin{equation}
\label{MRLS-eqROM4}
\begin{array}{l}
d \red{x}_t =(\red{A} \red{x}_t + \red{B} u_t)d\xi + (\red{F} \red{x}_t + \red{G} u_t)d\W_t, \\[2mm]
\red{y}_t=\red{C} \red{x}_t,
\end{array}
\end{equation}
is a stochastic model in the mean-square of system~(\ref{ORS-eq1}) at $(S,L)$.
\end{proposition}

\textit{Proof.} Under certain stability properties (that will be established later in the proof), the steady state of $\red{m}=E[\red{x}_t]$ (\textit{i.e.} of the mean of $\red{x}_t$) is $R\omega$, with $R$ the unique solution of  
\begin{equation}
\label{MRLS-eqmROM}
\red{A} R - R S = -\red{B} L.
\end{equation}
Similarly, the steady state of $\vect{\red{M}}=\vect{E[\red{x}_t\red{x}_t^{\top}]}$ is $\red{\mathcal{K}} \vect{\omega \omega^{\top}}$, where $\red{\mathcal{K}}$ is the unique solution of the equation
\begin{equation}
\label{MRLS-eqMROM}
\red{\mathcal{A}} \red{\mathcal{K}} + \red{\mathcal{B}}= \red{\mathcal{K}} \mathcal{S},
\end{equation}
with $\red{\mathcal{A}}=I\otimes \red{A} + \red{A}\otimes I + \red{F} \otimes \red{F}$, $\red{\mathcal{B}}=\red{B}L \otimes R+ R \otimes \red{B}L + \red{G}L \otimes \red{F} R + \red{F} R \otimes \red{G}L + \red{G}L \otimes \red{G}L$. Note that the mean-square of the output can be written as $\vect{E[y_ty_t^{\top}]}=\vect{CE[x_tx_t^{\top}]C^\top}=(C \otimes C)\vect{E[x_tx_t^{\top}]}$.
For system~(\ref{ORS-eq1}) and model~(\ref{MRLS-eqROM4}) to have the same steady-state mean-square of the output, we need that $(\red{C} \otimes \red{C})\red{\mathcal{K}} = (C \otimes C)\mathcal{K}$. This is achieved by setting $\red{\mathcal{K}}=I$ and $\red{C}$ such that $\red{C} \otimes \red{C} = \left( C \otimes C \right) \mathcal{K}$. 
Determine now any invertible matrix $R$ such that $\red{C}R=C\Pi$. Note that since $\red{C}$ is not identically zero, it is always possible to find an invertible matrix $R$ solving this equation. Now select $\red{A}=RSR^{-1}-\red{B}LR^{-1}$ and note that this selection solves equation~(\ref{MRLS-eqmROM}). The solution $R$ is unique for any matrix $\red{B}$ such that $\sigma(\red{A})\cap \sigma(S)=\emptyset$. This last condition is guaranteed if $\red{B}$ is selected such that $\sigma(\red{A})\subset \mathbb{C}_{<0}$. Moreover, this last property also ensures that the steady state of the mean of the output of (\ref{MRLS-eqROM4}) is well-defined and equal to $\red{C}R\omega$, which by construction is equal to $C\Pi\omega$. Hence, system~(\ref{ORS-eq1}) and model~(\ref{MRLS-eqROM4}) have the same steady-state output mean. Now note that if $\red{F}$ and $\red{G}$ are selected such that~(\ref{MRLS-eqFrGr}) holds, then equation~(\ref{MRLS-eqMROM}) has the unique solution $\red{\mathcal{K}}=I$. If in addition $\red{F}$ is such that $\textstyle\sigma(\red{\mathcal{A}})\subset\mathbb{C}_{<0}$, then the steady state of the mean-square of the output of~(\ref{MRLS-eqROM4}) is $(\red{C} \otimes \red{C})\red{\mathcal{K}} \vect{\omega \omega^{\top}}$, which is equal by construction to $(C \otimes C)\mathcal{K} \vect{\omega \omega^{\top}}$. Hence, system~(\ref{ORS-eq1}) and model~(\ref{MRLS-eqROM4}) have also the same steady-state output mean-square. \hspace*{0pt}\hfill $\square$ \par

In Proposition~\ref{MRLS-lmROM4} we need to determine $\red{C}$ from condition~(\ref{MRSS-eqC1M}). This problem is known as nearest Kronecker product approximation, which can be formulated as follows. Given a matrix $Q\in\mathbb{R}^{N\times N}$, the problem consists in determining the two matrices $T_1\in\mathbb{R}^{n_1}$ and $T_2\in\mathbb{R}^{n_2}$ such that $||Q - T_1 \otimes T_2 ||$ is minimized. The solution of this problem is given in \cite{VaLPit:93,Gen:07}. After rearranging the elements of $Q$ in a new matrix called $\bar{Q}\in\mathbb{R}^{n_1^2\times n_2^2}$ (see \cite{Gen:07} for details), we compute the singular value decomposition $U^{\top} \bar{Q}V=\diag(\delta_1,\dots,\delta_q)$ of $\bar{Q}$, with $q=\rank(\bar{Q})$. The solution to the problem is given by $\vect{T_1}=\sqrt{\delta_1}u_1$ and $\vect{T_2}=\sqrt{\delta_1}v_1$, where $u_1$ and $v_1$ are the first columns of $U$ and $V$, respectively. Note that the determination of the nearest Kronecker approximation introduces an error called separability approximation error, see \cite{Gen:07}. This error is zero if the rearranged matrix has only one singular value different from zero.

\begin{remark}
The main difficulty in the determination of the family of models~(\ref{MRLS-eqROM4}) is to solve equation~(\ref{MRLS-eqFrGr}). However, since $\red{G}$ is a free parameter, we can use it to simplify the computation of such a solution. Note in fact that any $\red{G}$ such that the conditions in Proposition~\ref{MRLS-lmROM4} hold would anyway give a reduced-order model in the mean-square by moment matching according to the definition given. Thus, the free parameter $\red{G}$ can be used to achieve properties besides moment matching. In this specific remark, we use the free parameter to simplify the computation of a reduced-order model. Thus, select $\red{G}=0$. As a result, equation~(\ref{MRLS-eqFrGr}) becomes
\begin{equation}
\label{MRSS-simFF}
\red{F} \otimes \red{F}=-I\otimes \red{A} - \red{A}\otimes I+I\otimes S + S\otimes I - \red{B}L \otimes R- R \otimes \red{B}L,
\end{equation}
from which we can determine $\red{F}$ as the nearest Kronecker approximation of $\red{F} \otimes \red{F}$. Although this selection simplifies the computation, the obtained model is not necessarily the best (in a sense which has to be defined) among the models belonging to the family~(\ref{MRLS-eqROM4}).
\end{remark}

\begin{remark}
The determination of a reduced-order model in the mean-square requires two nearest Kronecker approximations: the first is needed to compute $\red{C}$ from (\ref{MRSS-eqC1M}) and the second to compute $\red{F}$ from (\ref{MRLS-eqFrGr}). Conditions (\ref{MRSS-eqC1M}) and (\ref{MRLS-eqFrGr}) can be satisfied without error only if the respective rearranged matrices have only one non-zero singular value. In general, most of the time, a separability approximation error, which can be computed by means of the singular values, will be introduced.
The separability approximation error of $\red{C}\otimes\red{C}$ can be influenced by selecting other matrices $L$ and $S$, whereas the separability approximation error of (\ref{MRLS-eqFrGr}) can also be influenced using the matrix $\red{B}$.
\end{remark}

In the following simple example we illustrate the different behaviours of stochastic models in the mean and stochastic models in the mean-square.

\begin{example}
\begin{figure}
\centering
\includegraphics[width=\columnwidth,height=6cm]{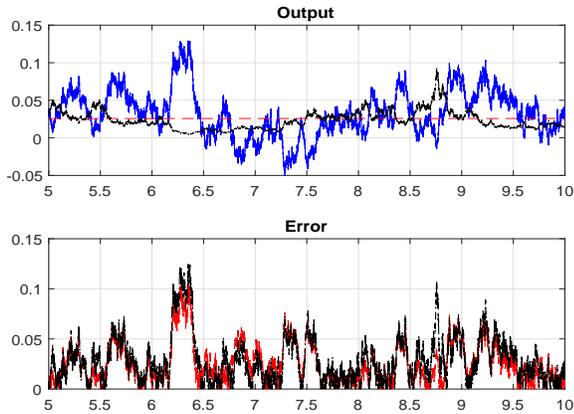}
\caption{Top graph: time history of the steady-state output of system~(\ref{ORS-eq1}) (solid/blue), of the output of the stochastic reduced-order model in the mean~(\ref{MRLS-eqROM3}) (dashed/red) and of the output of the stochastic reduced-order model in the mean-square~(\ref{MRLS-eqROM4}) (dash-dotted/black). Bottom graph: time history of the corresponding absolute errors.}
\label{MRSS-fig2}
\end{figure}%
\begin{figure}
\centering
\includegraphics[width=\columnwidth,height=6cm]{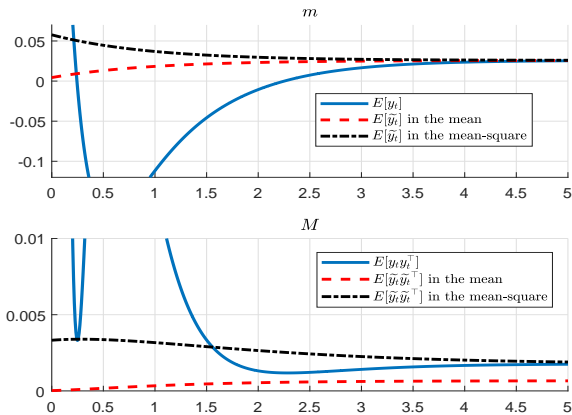}
\caption{Time history of the mean/mean-square (top/bottom) of output of system~(\ref{ORS-eq1}) (solid/blue), of the mean/mean-square (top/bottom) of output of the stochastic reduced-order model in the mean~(\ref{MRLS-eqROM3}) (dashed/red) and of the mean/mean-square (top/bottom) of the output of the stochastic reduced-order model in the mean-square~(\ref{MRLS-eqROM4}) (dotted/black).}
\label{MRSS-fig2mM}
\end{figure}%
Consider a linear system~(\ref{ORS-eq1}) of order $n=10$ with randomly generated matrices\footnote{All the matrices used in this simulation can be downloaded from \cite{FILE_MRSS}.} $A$, $B$ and $C$, and with $F=0.05A$ and $G=B$. Consider a signal generator~(\ref{ORS-eq2}) of order $\nu=1$ with $J$ zero, $S=0$ and $L$ randomly generated. A stochastic reduced-order model in the mean~(\ref{MRLS-eqROM3}) and a stochastic reduced-order model in the mean-square~(\ref{MRLS-eqROM4}) are computed. For both models we have selected $\red{B}$ such that the eigenvalue of $\red{A}$ is one of the eigenvalues of $A$. For the first model $\red{G}=0.1\red{B}$ and $\red{F}=-\red{G}L$. For the second model $\red{G}=0$ and $\red{F}$ is determined from~(\ref{MRSS-simFF}). The top graph in Fig.~\ref{MRSS-fig2} shows the steady-state output of the system (solid/blue), of the stochastic reduced-order model in the mean~(\ref{MRLS-eqROM3}) (dashed/red) and of the stochastic reduced-order model in the mean-square~(\ref{MRLS-eqROM4}) (dash-dotted/black). The bottom graph shows the corresponding absolute errors. 
One can assess the differences between the two reduced-order models by looking at the mean and mean-square of $y_t$. The top graph in Fig.~\ref{MRSS-fig2mM} shows the time history of $E[y_t]$ (solid/blue), of $E[\widetilde{y_t}]$ generated by the model in the mean (dashed/red) and of $E[\widetilde{y_t}]$ generated by the model in the mean-square (dotted/black). The bottom graph in Fig.~\ref{MRSS-fig2mM} shows the time history of the mean-square for the same models. We note that both reduced-order model preserve the mean at steady state, whereas only the second model preserves the mean-square at steady state.
\end{example}

In summary, the models in the moment-mean/mean~(\ref{MRLS-eqROM3}), the models in the mean-square~(\ref{MRLS-eqROM4}) and the stochastic models~(\ref{MRLS-eqROM2}) have increasing computational complexity but decreasing approximation error. From our discussion, it is also clear that models in the mean-square are also models in the mean. In fact, among the models preserving the mean, the models in the mean-square are the ones preserving \textit{also} the mean-square. 

\subsection{Algorithmic Discussion}
\label{MRSS-sec-AlgImp}

As already mentioned, the computation of the stochastic reduced-order model~(\ref{MRLS-eqROM2}) is costly. The cost of the reduced-order model in the moment-mean has complexity identical to classical deterministic model reduction methods and various state-of-the-art algorithms could be used to decrease the cost further. In fact, note that to determine model~(\ref{MRLS-eqROM3}) we just need to compute the vector $C\Pi$, where $\Pi$ is the solution of the Sylvester equation~(\ref{MRSS-eqEPI}). Thus one could proceed in a number of ``deterministic'' ways by defining an auxiliary deterministic system for which its steady state is described by the solution of the Sylvester equation~(\ref{MRSS-eqEPI}). When $J=0$ one could directly use the IRKA algorithm~\cite{GugAntBea:08}, which uses efficient Krylov projections, and then efficiently extract the matrix $C\Pi$ (since the obtained model is low dimension). Otherwise, one could use the auxiliary deterministic system to generate a trajectory and then apply the data-driven method presented in~\cite{ScaAst:16a}. In particular, this second method computes directly the matrix $C\Pi$ from data and it has a complexity of $\mathcal{O}(\alpha\nu)$, for some positive $\alpha$, in its most efficient form. When $J\ne0$ one could determine the solution of (\ref{MRSS-eqEPI}) with an efficient method of choice~\cite[Chapter 6]{Ant:05}.

The cost of the reduced-order model in the mean-square is larger, since in this case the matrix $\mathcal{K}$ has dimension $n^2 \times \nu^2$. Nevertheless, the computation of $\left( C \otimes C \right) \mathcal{K}$ can be done again using deterministic methods as equation~(\ref{MRLS-eqMSS}) is a standard Sylvester equation. Thus, the same remarks made for the models in the mean carry over to the model in the mean-square, keeping in mind that all complexities are squared. In addition, this family of models also requires two nearest Kronecker approximations and, consequently, two singular value decompositions.

\section{A benchmark system}
\label{MRSS-sec-Sim}

In this section we illustrate some of the results of the paper on a stochastic modification of a classical benchmark system, the Los Angeles University Hospital building model, used in the literature of deterministic model reduction. Although the model is relatively low dimensional ($n=48$), it presents several frequency response peaks that make it interesting. The deterministic model is described in \cite{ChaVaD:05} and the matrices can be downloaded from \cite{SLICOT}. Moreover, the model has been reduced to order $\nu=31$ with several deterministic techniques in \cite{Ant:05}. The model is described by a mechanical second-order differential equation 
$$
M_H \ddot q + C_H \dot q + K_H q = B_H u,
$$
where $q\in\mathbb{R}^{\kappa}$, $M_H\in\mathbb{R}^{\kappa \times \kappa}$, $C_H\in\mathbb{R}^{\kappa \times \kappa}$, $K_H\in\mathbb{R}^{\kappa \times \kappa}$ and $B_H\in\mathbb{R}^{\kappa \times  1}$, with $M_H$ positive definite. This system can be written in the form~(\ref{ORS-eq1}) with
$$
A=\left[\begin{array}{cc}0 & I\\ - M_H^{-1}K_H & -M_H^{-1}C_H\end{array}\right], \qquad B=\left[\begin{array}{c}0 \\ - M_H^{-1}B_H\end{array}\right].
$$
The matrix $C$ has all zero elements apart for its 25th element, which is equal to 1, corresponding to the displacement in the horizontal direction of the first floor of the building. The matrices $F$ and $G$ are selected as $F=0.01A$ and $G=B$. The matrix $S$ of the signal generator is selected as in \cite{ScaJiaAst:17,ScaAst:16a}, \textit{i.e.} a matrix of order $\nu=19$ with eigenvalues $0$, $\pm5.22\iota$, $\pm10.3\iota$, $\pm13.5\iota$, $\pm22.2\iota$, $\pm24.5\iota$, $\pm36\iota$, $\pm42.4\iota$, $\pm55.9\iota$ and $\pm70\iota$ (corresponding to the main frequency peaks of the deterministic model). The matrix $J$ is selected as the zero matrix.
\begin{figure}
\centering
\includegraphics[width=\columnwidth,height=6cm]{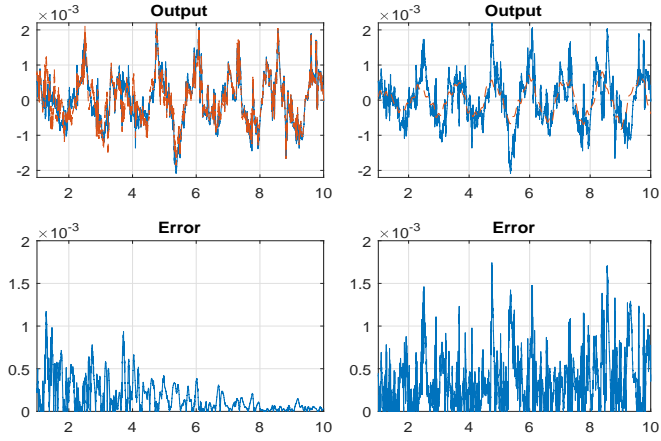}
\caption{Top left graph: time history of the output of system~(\ref{ORS-eq1}) (solid/blue) and of the output of the stochastic reduced-order model~(\ref{MRLS-eqROM2}) (dashed/red). Bottom left graph: time history of the corresponding absolute error. Top right graph: time history of the output of system~(\ref{ORS-eq1}) (solid/blue) and of the output of the stochastic reduced-order model in the mean~(\ref{MRLS-eqROM3}) (dashed/red). Bottom right graph: time history of the corresponding absolute error.}
\label{MRSS-fig3}
\end{figure}%
\begin{figure}
\centering
\includegraphics[width=\columnwidth,height=6cm]{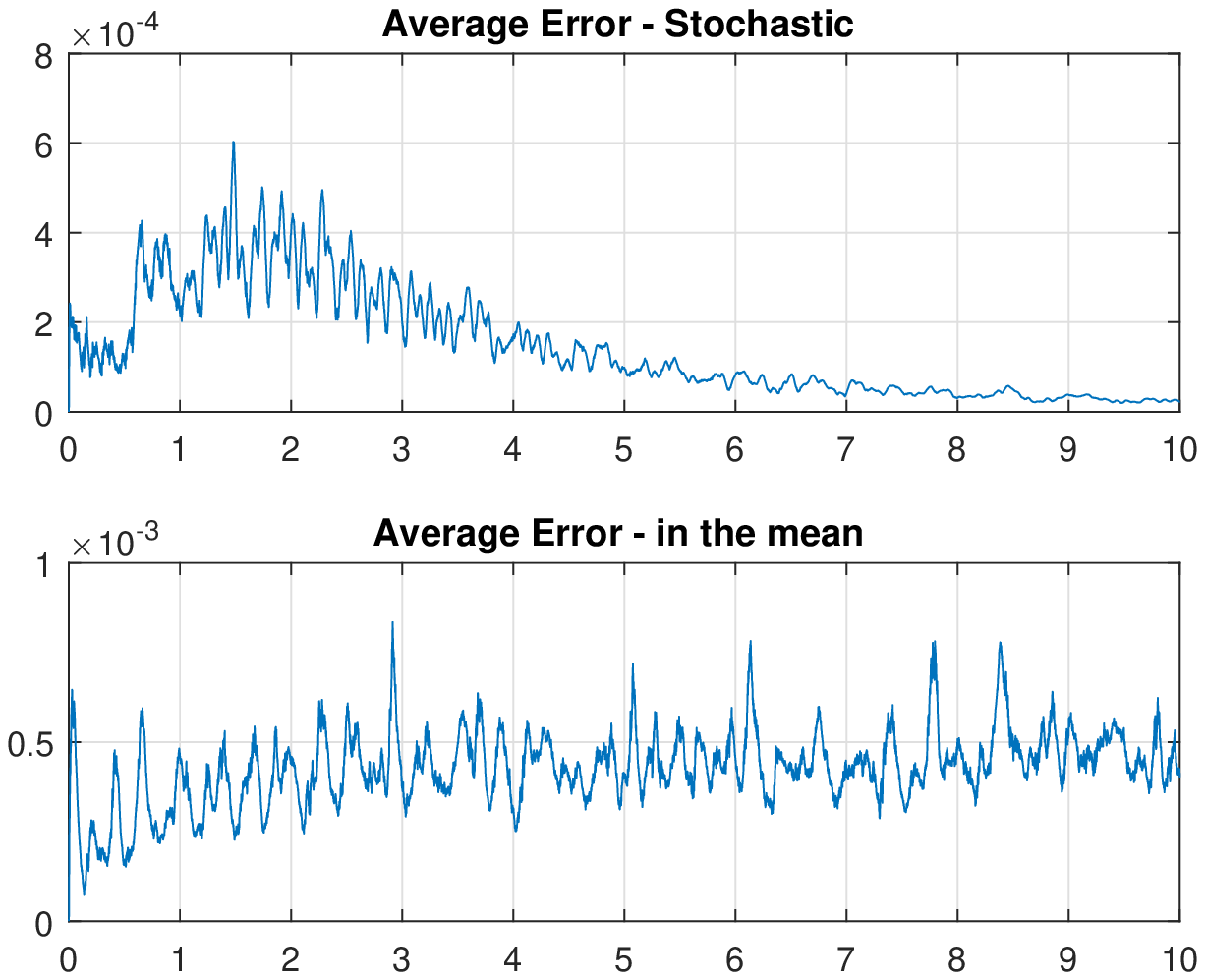}
\caption{Top graph: average of the time histories of the absolute error between the output of system~(\ref{ORS-eq1}) and of the output of the stochastic reduced-order model~(\ref{MRLS-eqROM2}) computed over 50 realizations (and with different $\omega_0$). Bottom graph: analogous, but with the error computed with the output of the stochastic reduced-order model in the mean~(\ref{MRLS-eqROM3}).}
\label{MRSS-fig3bis}
\end{figure}%
\begin{figure}
\centering
\includegraphics[width=\columnwidth,height=6cm]{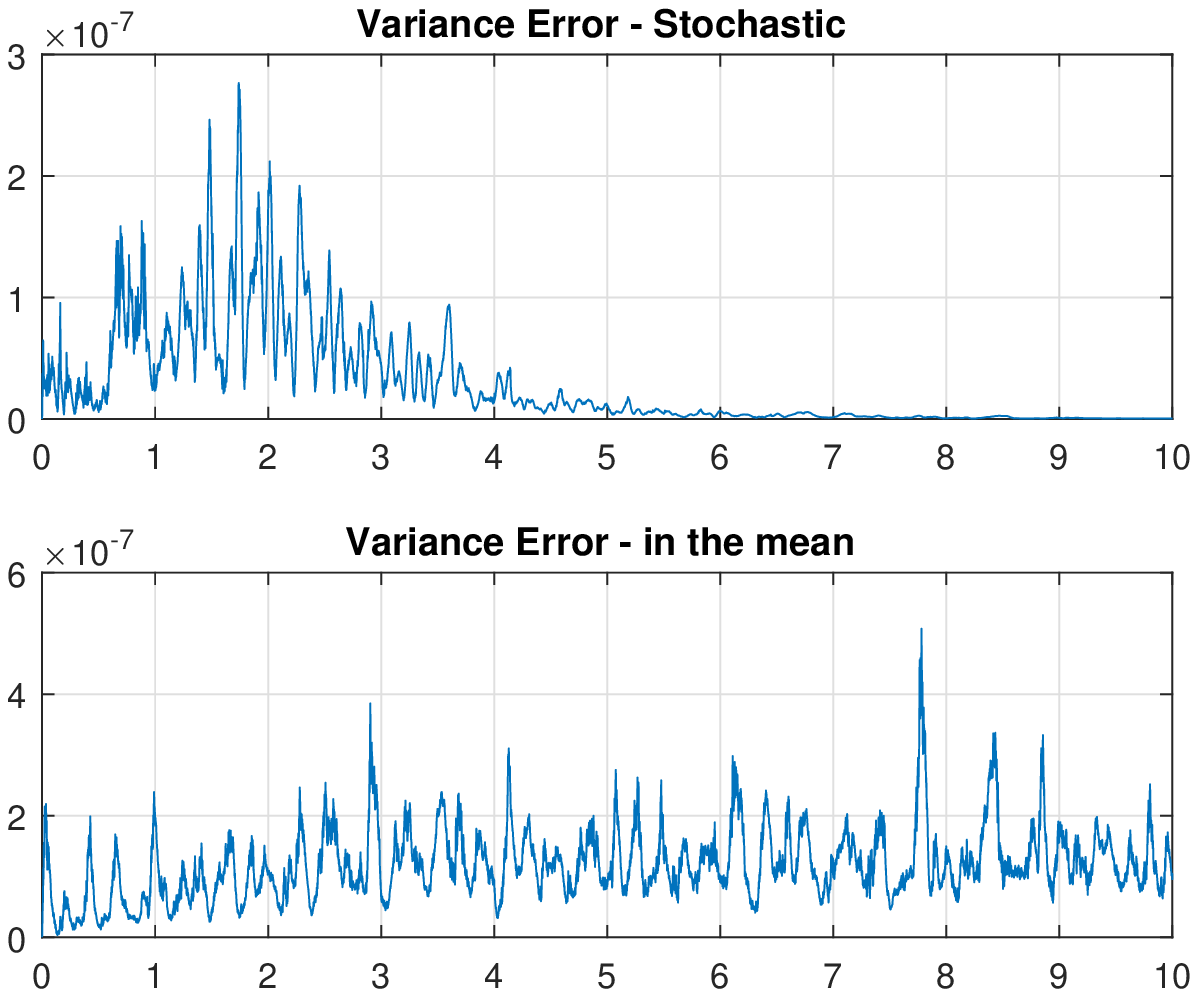}
\caption{Top graph: variance of the time histories of the absolute error between the output of system~(\ref{ORS-eq1}) and of the output of the stochastic reduced-order model~(\ref{MRLS-eqROM2}) computed over 50 realizations (and with different $\omega_0$). Bottom graph: analogous, but with the error computed with the output of the stochastic reduced-order model in the mean~(\ref{MRLS-eqROM3}).}
\label{MRSS-fig3tris}
\end{figure}%
\begin{figure}
\centering
\includegraphics[width=\columnwidth,height=6cm]{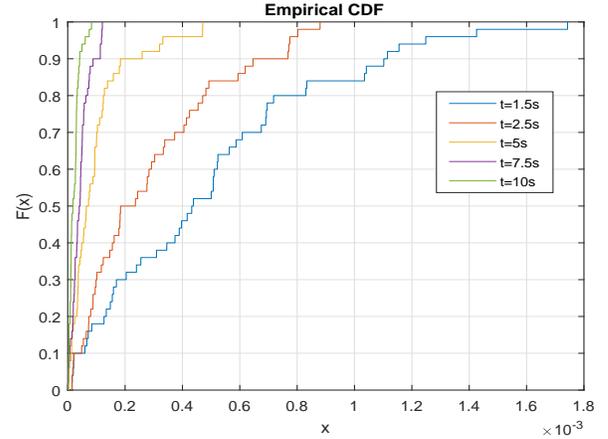}
\caption{Empirical cumulative distribution function computed from 50 realizations for the absolute error obtained with the stochastic reduced-order model~(\ref{MRLS-eqROM2}). The curves are parametrized with respect to time for $t=1.5$, $t=2.5$, $t=5$, $t=7.5$ and $t=10$ seconds.}
\label{MRSS-fig3quat}
\end{figure}%
\begin{figure}
\centering
\includegraphics[width=\columnwidth,height=6cm]{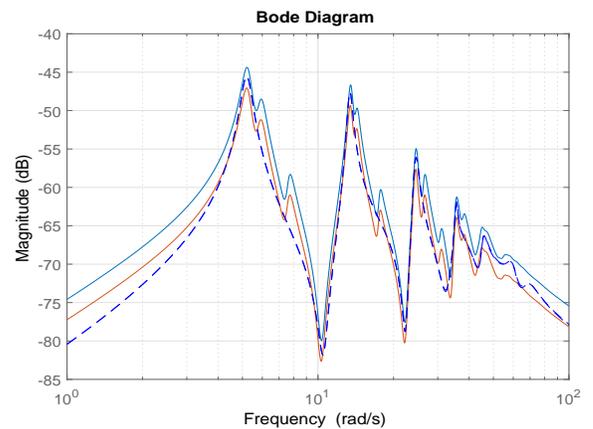}
\caption{Magnitude plot of system~(\ref{MRSS-eqExw1}) (solid lines) and of the model~(\ref{MRSS-eqExw2}) (dashed lines) for the two values $\bar{w}=\{10\max \Delta\W_t,\,10\min \Delta\W_t\}$.}
\label{MRSS-fig4}
\end{figure}%
\begin{figure}
\centering
\includegraphics[width=\columnwidth,height=6cm]{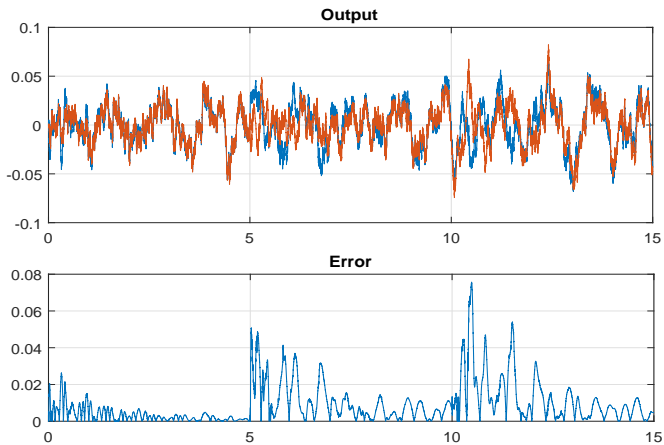}
\caption{Top graph: time history of the output of system~(\ref{ORS-eq1}) (solid/blue) and of the output of the stochastic reduced-order model~(\ref{MRLS-eqROM2}) (dashed/red) for the square input~(\ref{MRSS-eqSquare}). Bottom graph: time history of the corresponding absolute error.}
\label{MRSS-fig5}
\end{figure}%
A stochastic reduced-order model~(\ref{MRLS-eqROM2}) and a stochastic reduced-order model in the mean~(\ref{MRLS-eqROM3}) are computed\footnote{The matrices of the two reduced-order models can be downloaded from \cite{FILE_MRSS}.}. The two models have the same matrices $\red{A}$, $\red{B}$, $\red{F}$, $\red{H}$. The matrices $\red{A}$, $\red{B}$ and $R\ne I$ are selected using the method presented in \cite{ScaJiaAst:16,ScaJiaAst:17}. The matrices $\red{F}$ and $\red{G}$ are selected as $\red{G}=0.05\red{B}$ and $\red{F}=-\red{G}LR^{-1}$. For the first model the output mapping is $C \X_t R^{-1}$, whereas for the second model the output mapping is $C \Pi R^{-1}$. The output of the system is shown in solid/blue line in the top two graphs of Fig.~\ref{MRSS-fig3}. The left two graphs show also the output of the stochastic reduced-order model~(\ref{MRLS-eqROM2}) in dashed/red line (top) and the absolute error (bottom). The right two graphs show also the output of the stochastic reduced-order model in the mean~(\ref{MRLS-eqROM3}) in dashed/red line (top) and the absolute error (bottom). The behavior shown in the figure is consistent with the one seen in Fig.~\ref{MRSS-fig1}. Fig.~\ref{MRSS-fig3bis} shows the time history of the average of the absolute error $|y_t-\widetilde{y}_t|$ computed over 50 realizations (and every time with a different input generated by a different, randomly generated, $\omega_0$). In the top graph $\widetilde{y}_t$ is obtained from the stochastic reduced-order model~(\ref{MRLS-eqROM2}), whereas in the bottom graph $\widetilde{y}_t$ is obtained from the stochastic reduced-order model in the mean~(\ref{MRLS-eqROM3}). Fig.~\ref{MRSS-fig3tris} shows the variance of the same quantities. Fig.~\ref{MRSS-fig3bis} and \ref{MRSS-fig3tris} show that the results obtained in Fig.~\ref{MRSS-fig3} are consistent across realizations. Fig.~\ref{MRSS-fig3quat} shows the empirical cumulative distribution function computed from these 50 realizations for the absolute error obtained with the stochastic reduced-order model~(\ref{MRLS-eqROM2}), parametrized with respect to time (from $t=1.5$ to $t=10$ seconds). We see that as time increases, the distribution approaches zeros, \textit{i.e.} all realizations approach zero at steady state. To analyse the response of the models to non-interpolating signals (\textit{i.e.} not produced by the signal generator), we compare the Bode plot of the deterministic system
\begin{equation}
\label{MRSS-eqExw1}
\dot x = (A+F\bar{w})x+(B+G\bar{w})u, \qquad y=Cx,
\end{equation}
for the two values $\bar{w}=\{10\max \Delta\W_t,\,10\min \Delta\W_t\}$\footnote{The factor of 10 is for improving visibility in the figures. $\Delta\W_t$ represents the discrete variations of $\W_t$ in the simulation software.} and of the deterministic model
\begin{equation}
\label{MRSS-eqExw2}
\dot{\red{x}} = (\red{A}+\red{F}\bar{w})\red{x}+(\red{B}+\red{G}\bar{w})u, \qquad \red{y}=C\Pi R^{-1}\red{x},
\end{equation}
for the same values $\bar{w}$. The solid lines in Fig.~\ref{MRSS-fig4} show the magnitude plot of the two deterministic systems~(\ref{MRSS-eqExw1}), whereas the dashed lines (mostly overlapped) show the magnitude plot of the two deterministic reduced-order models~(\ref{MRSS-eqExw2}). By interpreting the stochastic system as a perturbed deterministic system, we notice that the deterministic part of the system plays a fundamental role in approximating the stochastic system, suggesting that classical deterministic moment matching techniques can be used to design this part of the stochastic reduced-order model. Finally, Fig.~\ref{MRSS-fig5} shows the output of system~(\ref{ORS-eq1}) (solid/blue) and of the output of the stochastic reduced-order model~(\ref{MRLS-eqROM2}) (dashed/red) for the square input
\begin{equation}
\label{MRSS-eqSquare}
u(t)= -0.05 \sign\left( \sin\left(\frac{2 \pi}{10} t\right)\right),
\end{equation}
which is not an input generated by the signal generator~(\ref{ORS-eq2}). We see that there is an error at the switching times but, as expected, the error decreases in the periods of time in which the input is constant (because 0 is an eigenvalue of $S$).

\section{Conclusions}
\label{MRSS-sec-Con}

In this paper we have studied the problem of model reduction by moment matching for linear and nonlinear stochastic systems. We have characterized the moment by means of a generalized Sylvester equation (for linear systems) and of a stochastic partial differential equation (for nonlinear systems) and we have then proposed families of reduced-order models. We have noticed that these models cannot be considered simpler and we have proposed various approximated models, based on different relaxations of the moment matching condition, which provide a computational advantage at the cost of introducing a steady-state error. We have reserved particular attention to linear systems and we have illustrated the results of the paper with several simulations.  An important future research direction consists in the development of efficient ways to compute the stochastic moment of the system, without approximations.


\appendix

We begin by introducing the notions of stochastic process and of Brownian motion.

\begin{definition}\cite[Section 1.8]{Arn:74}
A \textit{stochastic process} with state space $\mathbb{R}^n$ is a family $\{x_t,\,t\in\mathbb{R}\}$ of $\mathbb{R}^n$-valued random variables, \textit{i.e.} for every fixed $t\in\mathbb{R}$, $x_t(\cdot)$ is an $\mathbb{R}^n$-valued random variable and, for every fixed $w \in \Omega$, $x_{\cdot}(w)$ is an $\mathbb{R}^n$-valued function of time.
\end{definition}

\begin{definition}\label{MRSS-defB}\cite[Definition 1.1]{RogWil:94}
A stochastic process $\{\W_t,\,t\in\mathbb{R}_{\ge 0}\}$ is a \textit{Brownian motion} if
\begin{enumerate}
\item $\W_0(w)=0$ for all $w \in \Omega$;
\item the mapping $t\mapsto \W_t(w)$ is a continuous function of $t\in\R_{\ge 0}$ for all $w  \in \Omega$;
\item for every $t,\tau \ge 0$, $\W_{t+\tau}-\W_t$ is independent of $\{\W_{\bar{t}},\,0\le \bar{t} \le t\}$ and has Gaussian distribution with mean 0 and variance $\tau$.
\end{enumerate}
\end{definition}

All stochastic processes in the paper are adapted to the filtration $\mathcal{F}_t$ generated by the Brownian motion $\W_t$ up to time $t$, and possibly to the filtrations generated by other Brownian motions whenever they appear. This is a consequence of the theorem of existence and uniqueness of solutions of stochastic differential equations, see \textit{e.g.}  \cite[Theorem 5.2.1]{Oks:13}.
We now briefly recall some formulas and properties which are instrumental to derive the results of the paper. First of all, recall that the relations
\begin{equation}
\label{ORS-eqIto1}
(dt)^2=dtd\W_t=0, 
\end{equation}
and 
\begin{equation}
\label{ORS-eqIto2}
\qquad (d\W_t)^2=dt,
\end{equation}
hold for the differential of a stochastic process (see \textit{e.g.} \cite[Chapter 2]{Gar:88} or \cite[Chapter 4]{Oks:13}). Given a function $p:(t,\omega_t)\mapsto p(t,\omega_t)$, It\^{o}'s Lemma provides an explicit formula for its differential. We produce here a straightforward extension of It\^{o}'s Lemma applied to $p(t,\W_t,\omega_t)$.

\begin{lemma}
\label{MRSS-lmIto}(It\^{o}'s formula)
Let $p : \R \times \R \times \R^{\nu} \to \R^{\nu} : (t,W_t,\omega_t) \mapsto p(t,W_t,\omega_t)$, with $\omega_t$ given by (\ref{MRNS-eq2}), be a continuous function with continuous partial derivatives. Then the stochastic process $\chi_t=p(t,\W_t,\omega_t)$ possesses a stochastic differential given by
$$
\begin{array}{rl}
d\chi_t&\!\!\!\!=\left[\frac{\partial p}{\partial t}+ \frac{\partial p}{\partial \omega_t}s(\omega_t) + \frac{1}{2} \frac{\partial^2 p}{\partial \W_t^2} \right. + \frac{\partial^2 p}{\partial \W_t \partial \omega_t}j(\omega_t)\\
&\!\!\!\!\left.+\frac{1}{2} j(\omega_t)^{\top}\frac{\partial^2 p}{\partial \omega_t^2} j(\omega_t) \right]dt + \left[\frac{\partial p}{\partial \W_t} + \frac{\partial p}{\partial \omega_t}j(\omega_t) \right]d\W_t.
\end{array}
$$
\end{lemma}

\textit{Proof:}
The formula follows by repeating the steps in \cite[Proof of Theorem 2.10]{Gar:88} or \cite[Proof of Theorem 4.1.2]{Oks:13} on the Taylor expansion of the function $p(t,\W_t,\omega_t)$ and using the relations (\ref{ORS-eqIto1}) and (\ref{ORS-eqIto2}) to cancel the higher order terms.
\hspace*{0pt}\hfill $\square$ \par

By using the integration by parts formula \cite[Theorem 4.1.5]{Oks:13}, it is easy to derive the following formula which is often used in the paper.

\begin{lemma}\label{MRSS-lmdd}(Stochastic product rule)\cite[Exercise 4.3]{Oks:13}
Let $a_t$ and $b_t$ be two stochastic processes. Then $d(a_tb_t) = da_t b_t + a_t db_t + da_tdb_t$.
\end{lemma}

We now recall the notion of stability for stochastic systems which is used in the paper.


We say that an event $\varepsilon\in\mathcal{F}$ happens \emph{almost surely} if $\mathcal{P}(\varepsilon) = 1$, see \emph{e.g.} \cite{Oks:13}.

\begin{definition}
    \cite{Koz:69} The equilibrium $x_t \equiv 0$ of system~(\ref{MRNS-eq1}) is said to be almost surely stable if
    \begin{equation}
        \mathcal{P}\left(\lim_{\Vert x_{t_0}\Vert\rightarrow 0}\sup_{t>t_0}\Vert x_t\Vert = 0 \right) = 1.
    \end{equation}
    The equilibrium $x_t \equiv 0$ of system~(\ref{MRNS-eq1}) is said to be almost surely asymptotically stable if it is almost surely stable and there exists $\delta'>0$ such that $\Vert x_{t_0}\Vert < \delta'$ implies for any $\epsilon>0$
    \begin{equation}
        \lim_{\delta\rightarrow \infty}\mathcal{P}\left(\sup_{t>\delta}\Vert x_t \Vert >\epsilon \right) = 0.
    \end{equation}
The zero equilibrium of system~(\ref{MRNS-eq1}) is \textit{almost surely asymptotically unbounded\footnote{\cite{Kha:11} uses the word ``unstable'' instead of ``unbounded''.}} (see \cite[Theorem 6.11]{Kha:11}) if $\p\left\{\lim_{t\to\infty}|x_t|=\infty\right\}=1$.
\end{definition}

A useful tool to assess the stability properties of a stochastic systems is the Lyapunov spectrum.

\begin{definition}\cite[Chapters 6.7 and 6.8]{Kha:11}
Consider system~(\ref{ORS-eq1}) and its fundamental matrix $\Phi_t$. Let $v_i$, with $i=1,\dots, n$, be $n$ linearly independent vectors. The \textit{Lyapunov exponent} of $\Phi_t$ in the direction $v_i\in\R^n$ is defined as $\textstyle \lambda_i = \limsup_{t\to +\infty} \frac{1}{t} \log ||\Phi_t v_i||$. The set of all Lyapunov exponents $\{\lambda_i$, $i=1,\dots,n\}$ of $\Phi_t$ is called \textit{Lyapunov spectrum} of $\Phi_t$ which we indicate with the symbol $\sigma_L(\Phi_t)$.
\end{definition}

\begin{theorem}\label{thm-stab}\cite[Theorem 6.11 and 6.12]{Kha:11}
Consider system~(\ref{ORS-eq1}) and its fundamental matrix $\Phi_t$.
If all Lyapunov exponents of $\Phi_t$ are negative, then system~(\ref{ORS-eq1}) is almost surely asymptotically stable. If at least one Lyapunov exponent of $\Phi_t$ is positive, then system~(\ref{ORS-eq1}) is almost surely asymptotically unbounded. If all Lyapunov exponents of $\Phi_t$ are zero then system~(\ref{ORS-eq1}) is neither almost surely asymptotically stable nor almost surely asymptotically unbounded.
\end{theorem}

Note that the conditions of Theorem~\ref{thm-stab} can be verified using the characterization of Lyapunov exponent given just above [62, Theorem 6.11] (therein called $a^*$). A worked example of this is shown in [62, Section 6.9].\\
To clarify the meaning of the last statement of Theorem~\ref{thm-stab}, we consider a simple example.

\begin{example}
If $S$ and $J$ commute and all the eigenvalues of $\textstyle S-\frac{1}{2}J^2$ are simple and have zero real part, then all the Lyapunov exponents of $\Sigma_t$ are zero. In the deterministic case ($J=0$) this implies simple stability and boundedness of trajectories. In the stochastic case ($J\ne 0$) neither of these properties carry over \cite[p. 140]{Gar:88}. 
\end{example}

Simple stability and boundedness are classically associated to the center manifold theory, which is the tool used to characterize steady-state solutions in the deterministic framework. Boxler showed in \cite{Box:89} that, although stability and boundedness are not provided by zero Lyapunov exponents, these are the correct objects to characterize stochastic center manifolds. To show this, suppose that the linearization of system~(\ref{MRNS-eq1}) around zero has $n_s$ negative Lyapunov exponents, $n_u$ positive Lyapunov exponents and $n_c$ zero Lyapunov exponents. Then the state-space $\R^n$ of (\ref{MRNS-eq1}) can be decomposed as
$$
\R^n= E_c(w) \oplus E_s(w) \oplus E_u(w)
$$
where $E_c$, $E_s$ and $E_u$, which are called  Oseledec spaces, are associated to the zero, negative and positive Lyapunov exponents, respectively (see \cite[Sections 2 and 4]{Box:89} for a complete characterization of these spaces). Let\footnote{If $n_s=0$, then $\lambda_s=-\infty$. If $n_u=0$, then $\lambda_u=+\infty$.}
$$
\lambda_s:=\!\! \max_{\displaystyle\lambda_i\in\sigma_L(\Phi_t)\cap\R_{<0}}\!\! \lambda_i, \quad \lambda_u:= \!\!\min_{\displaystyle\lambda_i\in\sigma_L(\Phi_t)\cap\R_{>0}}\!\! \lambda_i.
$$
Below, after recalling the definition of the random norm given in \cite{Box:89}, we recall the characterization of the center manifold by means of its dynamical properties.
\begin{definition}
\label{def-rn}(\cite[Lemma 4.2]{Box:89})
Let $\beta>0$ be given such that $\lambda_s+4\beta<0$ and $\lambda_u-4\beta>0$. Then we define the random norm $|\cdot|_{w}$ as
$$
\!\!\!\!\begin{array}{ll}
|x|_w := \int_0^\infty e^{-(\lambda_s + 2 \beta)\tau}||\Phi^s(\tau,w)x||d\tau, & \!\!\text{for any }x\in E_s(w),\\[2mm]
|x|_w := \int_{-\infty}^\infty e^{- 2 \beta|\tau|}||\Phi^c(\tau,w)x||d\tau, & \!\!\text{for any }x\in E_c(w),\\[2mm]
|x|_w := \int_0^\infty e^{(\lambda_s - 2 \beta)\tau}||\Phi^u(-\tau,w)x||d\tau, & \!\!\text{for any }x\in E_u(w),\\[2mm]
\end{array}
$$
where $\Phi^s_t$, $\Phi^c_t$ and $\Phi^u_t$ are the sub-blocks of the fundamental matrix $\Phi_t$ corresponding to $E_s$, $E_c$ and $E_u$, respectively.   
\end{definition}
\begin{theorem}\cite[Section 7.2]{Box:89}
\label{MRSS-thmBox}
Let $\beta>0$ be such that $\lambda_s+4\beta<0$ and $\lambda_u-4\beta>0$. Then for any $\delta$ such that $2\beta < \delta < \min(-(\lambda_s+2\beta),\lambda_u-2\beta)$ the random set
$$
\begin{array}{l}
W_{\text{dyn}} := \left\{x_t: \limsup_{t \to + \infty} \frac{1}{t} \log | x_t |_{w} \le \delta \text{ and }\right. \\
\qquad\qquad\qquad\qquad\qquad\qquad\quad \left.\limsup_{t \to - \infty} \frac{1}{t} \log | x_t |_{w} \ge -\delta  \right\}
\end{array}
$$
is the stochastic center manifold, \textit{i.e.} $E_c(w)$ is the tangent space to $W_{\text{dyn}}$ at zero.
\end{theorem}

Hence, differently from the deterministic case, the steady-state solution of a system is not necessarily bounded forward and backward in time, but rather it satisfies the dynamic behaviour described in Theorem~\ref{MRSS-thmBox}.

\begin{remark}
According to Definition~\ref{MRSS-defB}, $\W_t$ is defined only for non-negative times. Since we want to characterize steady-state solutions in both directions of time (\textit{e.g.} as done in Theorem~\ref{MRSS-thmBox}) we consider extended Brownian motions made by joining at $t=0$ two independent copies of a Brownian motion, one defined for $t\in\R_{\ge 0}$ and one defined for $t\in\R_{\le 0}$. See \cite[Section 3.1]{Box:89} for more detail. 
\end{remark}


\bibliographystyle{IEEEtran}
\bibliography{bibdb}

\newcommand{\ACC}[1]{Proceedings of the #1 American Control
  Conference}\newcommand{\ECC}[1]{Proceedings of the #1 European Control
  Conference}\newcommand{\CDC}[1]{Proceedings of the #1 IEEE Conference on
  Decision and Control}
\begin{thebibliography}{10}
\providecommand{\url}[1]{#1}
\csname url@samestyle\endcsname
\providecommand{\newblock}{\relax}
\providecommand{\bibinfo}[2]{#2}
\providecommand{\BIBentrySTDinterwordspacing}{\spaceskip=0pt\relax}
\providecommand{\BIBentryALTinterwordstretchfactor}{4}
\providecommand{\BIBentryALTinterwordspacing}{\spaceskip=\fontdimen2\font plus
\BIBentryALTinterwordstretchfactor\fontdimen3\font minus
  \fontdimen4\font\relax}
\providecommand{\BIBforeignlanguage}[2]{{%
\expandafter\ifx\csname l@#1\endcsname\relax
\typeout{** WARNING: IEEEtran.bst: No hyphenation pattern has been}%
\typeout{** loaded for the language `#1'. Using the pattern for}%
\typeout{** the default language instead.}%
\else
\language=\csname l@#1\endcsname
\fi
#2}}
\providecommand{\BIBdecl}{\relax}
\BIBdecl

\bibitem{KarShr:91}
I.~Karatzas and S.~E. Shreve, \emph{Brownian Motion and Stochastic Calculus},
  ser. Graduate Texts in Mathematics.\hskip 1em plus 0.5em minus 0.4em\relax
  Springer New York, 1991.

\bibitem{YonZho:99}
J.~Yong and X.~Y. Zhou, \emph{Stochastic Controls: {H}amiltonian Systems and
  {HJB} Equations}, ser. Stochastic Modelling and Applied Probability.\hskip
  1em plus 0.5em minus 0.4em\relax Springer New York, 1999.

\bibitem{Mun:11}
C.~Munk, \emph{Fixed Income Modelling}.\hskip 1em plus 0.5em minus 0.4em\relax
  OUP Oxford, 2011.

\bibitem{Oks:13}
B.~{\O}ksendal, \emph{Stochastic Differential Equations: An Introduction with
  Applications}, 6th~ed., ser. Universitext.\hskip 1em plus 0.5em minus
  0.4em\relax Springer, 2013.

\bibitem{Kal:60}
R.~E. Kalman, ``A new approach to linear filtering and prediction problems,''
  \emph{ASME Journal of Basic Engineering}, vol.~82, no.~1, pp. 35--45, 1960.

\bibitem{KalBuc:61}
R.~E. Kalman and R.~S. Bucy, ``New results in linear filtering and prediction
  theory,'' \emph{ASME Journal of Basic Engineering}, pp. 95--108, March 1961.

\bibitem{LinPic:15}
A.~Lindquist and G.~Picci, \emph{Linear Stochastic Systems: A Geometric
  Approach to Modeling, Estimation and Identification}, ser. Series in
  Contemporary Mathematics.\hskip 1em plus 0.5em minus 0.4em\relax Springer
  Berlin Heidelberg, 2015.

\bibitem{Ant:05}
A.~Antoulas, \emph{Approximation of Large-Scale Dynamical Systems}.\hskip 1em
  plus 0.5em minus 0.4em\relax Philadelphia, PA: SIAM Advances in Design and
  Control, 2005.

\bibitem{ScaAst:17}
G.~Scarciotti and A.~Astolfi, ``Nonlinear model reduction by moment matching,''
  \emph{Foundations and Trends in Systems and Control}, vol.~4, no. 3-4, pp.
  224--409, 2017.

\bibitem{FaeScaAstRin:18}
N.~Faedo, G.~Scarciotti, A.~Astolfi, and J.~Ringwood, ``Energy-maximising
  control of wave energy converters using a moment-domain representation,''
  \emph{Control Engineering Practice}, vol.~81, pp. 85--96, 2018.

\bibitem{BreForScaAst:19}
V.~Breschi, S.~Formentin, G.~Scarciotti, and A.~Astolfi, ``Simulation-driven
  fixed-order controller tuning via moment matching,'' in \emph{\ECC{2019}},
  2019, pp. 2307--2312.

\bibitem{FaeScaAstRin:21}
N.~Faedo, G.~Scarciotti, A.~Astolfi, and J.~Ringwood, ``Nonlinear
  energy-maximising optimal control of wave energy systems: A moment-based
  approach,'' \emph{To appear in IEEE Transactions on Control Systems
  Technology}, 2021.

\bibitem{DesPal:84}
U.~B. Desai and D.~Pal, ``A transformation approach to stochastic model
  reduction,'' \emph{IEEE Transactions on Automatic Control}, vol.~29, no.~12,
  pp. 1097--1100, Dec 1984.

\bibitem{DesPalKir:85}
U.~B. Desai, D.~Pal, and R.~D. Kirkpatrick, ``A realization approach to
  stochastic model reduction,'' \emph{International Journal of Control},
  vol.~42, no.~4, pp. 821--838, 1985.

\bibitem{Tug:85}
J.~K. Tugnait, ``Order reduction of {SISO} nonminimum phase stochastic
  systems,'' in \emph{\CDC{24th}, Fort Lauderdale, FL, USA}, Dec 1985, pp.
  407--412.

\bibitem{Gre:88}
M.~Green, ``Balanced stochastic realizations,'' \emph{Linear Algebra and its
  Applications}, vol.~98, pp. 211--247, 1988.

\bibitem{Gre:88a}
------, ``A relative error bound for balanced stochastic truncation,''
  \emph{IEEE Transactions on Automatic Control}, vol.~33, no.~10, pp. 961--965,
  Oct 1988.

\bibitem{HarJonSil:84}
P.~Harshavardhana, E.~Jonckheere, and L.~Silverman, ``Stochastic balancing and
  approximation-stability and minimality,'' \emph{IEEE Transactions on
  Automatic Control}, vol.~29, no.~8, pp. 744--746, Aug 1984.

\bibitem{WanSaf:90}
W.~Wang and M.~G. Safonov, ``A tighter relative-error bound for balanced
  stochastic truncation,'' \emph{Systems \& Control Letters}, vol.~14, no.~4,
  pp. 307--317, 1990.

\bibitem{WanSaf:91}
------, ``Relative-error bound for discrete balanced stochastic truncation,''
  \emph{International Journal of Control}, vol.~54, no.~3, pp. 593--612, 1991.

\bibitem{LinPic:96}
A.~Lindquist and G.~Picci, ``Canonical correlation analysis, approximate
  covariance extension, and identification of stationary time series,''
  \emph{Automatica}, vol.~32, no.~5, pp. 709--733, 1996.

\bibitem{XuChe:03}
S.~Xu and T.~Chen, ``{$\mathcal{H}_{\infty}$} model reduction in the stochastic
  framework,'' \emph{SIAM Journal on Control and Optimization}, vol.~42, no.~4,
  pp. 1293--1309, 2003.

\bibitem{SuWuShiSon:12}
X.~Su, L.~Wu, P.~Shi, and Y.~D. Song, ``$\mathcal{H}_{\infty}$ model reduction
  of {T}akagi-{S}ugeno fuzzy stochastic systems,'' \emph{IEEE Transactions on
  Systems, Man, and Cybernetics, Part B (Cybernetics)}, vol.~42, no.~6, pp.
  1574--1585, Dec 2012.

\bibitem{BenDam:11}
P.~Benner and T.~Damm, ``{L}yapunov equations, energy functionals, and model
  order reduction of bilinear and stochastic systems,'' \emph{SIAM Journal on
  Control and Optimization}, vol.~49, no.~2, pp. 686--711, 2011.

\bibitem{BenRed:15}
P.~Benner and M.~Redmann, ``Model reduction for stochastic systems,''
  \emph{Stochastic Partial Differential Equations: Analysis and Computations},
  vol.~3, no.~3, pp. 291--338, 2015.

\bibitem{SchHak:86}
G.~Sch{\"o}ner and H.~Haken, ``The slaving principle for {S}tratonovich
  stochastic differential equations,'' \emph{Zeitschrift f{\"u}r Physik {B}
  Condensed Matter}, vol.~63, no.~4, pp. 493--504, Dec 1986.

\bibitem{SchHak:87}
------, ``A systematic elimination procedure for {I}to stochastic differential
  equations and the adiabatic approximation,'' \emph{Zeitschrift f{\"u}r Physik
  {B} Condensed Matter}, vol.~68, no.~1, pp. 89--103, Mar 1987.

\bibitem{XuRob:96}
C.~Xu and A.~Roberts, ``On the low-dimensional modelling of {S}tratonovich
  stochastic differential equations,'' \emph{Physica {A}: Statistical Mechanics
  and its Applications}, vol. 225, no.~1, pp. 62--80, 1996.

\bibitem{Rob:08}
A.~Roberts, ``Normal form transforms separate slow and fast modes in stochastic
  dynamical systems,'' \emph{Physica {A}: Statistical Mechanics and its
  Applications}, vol. 387, no.~1, pp. 12--38, 2008.

\bibitem{Rob:14}
------, \emph{Model Emergent Dynamics in Complex Systems:}, ser. Mathematical
  Modeling and Computation.\hskip 1em plus 0.5em minus 0.4em\relax Society for
  Industrial and Applied Mathematics, 2014.

\bibitem{CoiKevLafMagNad:08}
R.~R. Coifman, I.~G. Kevrekidis, S.~Lafon, M.~Maggioni, and B.~Nadler,
  ``Diffusion maps, reduction coordinates, and low dimensional representation
  of stochastic systems,'' \emph{Multiscale Modeling \& Simulation}, vol.~7,
  no.~2, pp. 842--864, 2008.

\bibitem{Nur:14}
H.~I. Nurdin, ``Structures and transformations for model reduction of linear
  quantum stochastic systems,'' \emph{IEEE Transactions on Automatic Control},
  vol.~59, no.~9, pp. 2413--2425, Sept 2014.

\bibitem{TecNur:16}
O.~Techakesari and H.~I. Nurdin, ``Tangential interpolatory projection for
  model reduction of linear quantum stochastic systems,'' \emph{To appear on
  IEEE Transactions on Automatic Control}, 2016.

\bibitem{PelMunKha:06}
S.~Pele{\v s}, B.~Munsky, and M.~Khammash, ``Reduction and solution of the
  chemical master equation using time scale separation and finite state
  projection,'' \emph{The Journal of Chemical Physics}, vol. 125, no.~20, pp.
  204\,104(1--13), 2006.

\bibitem{SinHes:10}
A.~Singh and J.~P. Hespanha, ``Stochastic hybrid systems for studying
  biochemical processes,'' \emph{Philosophical Transactions of the Royal
  Society of London A: Mathematical, Physical and Engineering Sciences}, vol.
  368, no. 1930, pp. 4995--5011, 2010.

\bibitem{BruChaSmi:14}
M.~Bruna, S.~J. Chapman, and M.~J. Smith, ``Model reduction for slow-fast
  stochastic systems with metastable behaviour,'' \emph{The Journal of Chemical
  Physics}, vol. 140, no.~17, 2014.

\bibitem{SooAnd:14}
A.~Sootla and J.~Anderson, ``On projection-based model reduction of biochemical
  networks part {II}: The stochastic case,'' in \emph{\CDC{53rd}}, Dec 2014,
  pp. 3621--3626.

\bibitem{GupKha:14}
A.~Gupta and M.~Khammash, ``Sensitivity analysis for stochastic chemical
  reaction networks with multiple time-scales,'' \emph{Electronic Journal of
  Probability}, vol.~19, no.~59, pp. 1--53, 2014.

\bibitem{MelHesKha:14}
B.~M{\'e}lyk{\'u}ti, J.~P. Hespanha, and M.~Khammash, ``Equilibrium
  distributions of simple biochemical reaction systems for time-scale
  separation in stochastic reaction networks,'' \emph{Journal of The Royal
  Society Interface}, vol.~11, no.~97, 2014.

\bibitem{JohBarSejMjo:15}
T.~Johnson, T.~Bartol, T.~Sejnowski, and E.~Mjolsness, ``Model reduction for
  stochastic {CaMKII} reaction kinetics in synapses by graph-constrained
  correlation dynamics,'' \emph{Physical Biology}, vol.~12, no.~4, 2015.

\bibitem{SmiCiaGri:15}
S.~Smith, C.~Cianci, and R.~Grima, ``Model reduction for stochastic chemical
  systems with abundant species,'' \emph{The Journal of Chemical Physics}, vol.
  143, no.~21, pp. 214\,105(1--19), 2015.

\bibitem{Sch:93}
J.~M.~A. Scherpen, ``Balancing for nonlinear systems,'' \emph{Systems \&
  Control Letters}, vol.~21, no.~2, pp. 143--153, Aug 1993.

\bibitem{Ast:10}
A.~Astolfi, ``Model reduction by moment matching for linear and nonlinear
  systems,'' \emph{IEEE Transactions on Automatic Control}, vol.~55, no.~10,
  pp. 2321--2336, 2010.

\bibitem{ChaVaD:05}
Y.~Chahlaoui and P.~{Van Dooren}, \emph{Dimension Reduction of Large-Scale
  Systems: Proceedings of a Workshop held in Oberwolfach, Germany, October
  19-25, 2003}.\hskip 1em plus 0.5em minus 0.4em\relax Berlin, Heidelberg:
  Springer, 2005, ch. Benchmark Examples for Model Reduction of Linear
  Time-Invariant Dynamical Systems, pp. 379--392.

\bibitem{SLICOT}
{SLICOT}, ``Benchmark examples for model reduction,''
  \url{http://slicot.org/20-site/126-benchmark-examples-for-model-reduction},
  accessed: 2018-06-07.

\bibitem{ScaTee:17}
G.~Scarciotti and A.~R. Teel, ``Model order reduction of stochastic linear
  systems by moment matching,'' in \emph{20th IFAC World Congress, Toulouse,
  France, July 9-14}, 2017, pp. 6506--6511.

\bibitem{ScaTee:17a}
------, ``Model order reduction for stochastic nonlinear systems,'' in
  \emph{\CDC{56th}, Melbourne, Australia, December 12-15}, 2017, pp.
  3069--3074.

\bibitem{Isi:95}
A.~Isidori, \emph{Nonlinear Control Systems}, {T}hird~ed., ser. Communications
  and Control Engineering.\hskip 1em plus 0.5em minus 0.4em\relax Springer,
  1995.

\bibitem{Sca:18}
G.~Scarciotti, ``Output regulation of linear stochastic systems: The
  full-information case,'' in \emph{2018 European Control Conference, Cyprus,
  June 12-15}, 2018, pp. 1920--1925.

\bibitem{MelSca:19}
A.~Mellone and G.~Scarciotti, ``{$\varepsilon$-Approximate Output Regulation of
  Linear Stochastic Systems: a Hybrid Approach},'' in \emph{2019 European
  Control Conference (ECC)}, June 2019, pp. 287--292.

\bibitem{MelSca:19a}
------, ``{Error-Feedback Output Regulation of Linear Stochastic Systems: a
  Hybrid Nonlinear Approach},'' in \emph{Joint Conference 8th IFAC Symposium on
  Mechatronic Systems (MECHATRONICS 2019), and 11th IFAC Symposium on Nonlinear
  Control Systems (NOLCOS 2019)}, September 2019, pp. 907--912.

\bibitem{MelSca:20}
------, ``Output regulation of linear stochastic systems,'' \emph{Conditionally
  accepted in IEEE Transactions on Automatic Control}, 2020.

\bibitem{Arn:74}
L.~Arnold, \emph{Stochastic differential equations}, ser. A Wiley-Interscience
  publication.\hskip 1em plus 0.5em minus 0.4em\relax Wiley, 1974.

\bibitem{Gar:88}
T.~C. Gard, \emph{Introduction to Stochastic Differential Equations}, ser.
  Monographs and textbooks in pure and applied mathematics.\hskip 1em plus
  0.5em minus 0.4em\relax M. Dekker, 1988.

\bibitem{Car:81}
J.~Carr, \emph{Applications of Centre Manifold Theory}, ser. Applied
  Mathematical Sciences Series.\hskip 1em plus 0.5em minus 0.4em\relax
  Springer-Verlag, 1981, no. v. 35.

\bibitem{Box:89}
P.~Boxler, ``A stochastic version of center manifold theory,''
  \emph{Probability Theory and Related Fields}, vol.~83, no.~4, pp. 509--545,
  1989.

\bibitem{ScaAst:15c}
G.~Scarciotti and A.~Astolfi, ``Model reduction by matching the steady-state
  response of explicit signal generators,'' \emph{IEEE Transactions on
  Automatic Control}, vol.~61, no.~7, pp. 1995--2000, 2016.

\bibitem{ScaAst:16d}
------, ``Model reduction for hybrid systems with state-dependent jumps,'' in
  \emph{IFAC Symposium Nonlinear Control Systems, Monterey, CA, USA}, 2016, pp.
  862--867.

\bibitem{ScaTeeAst:17}
G.~Scarciotti, A.~R. Teel, and A.~Astolfi, ``Model reduction for linear
  differential inclusions: moment-set and time-variance,'' in \emph{\ACC{2017},
  Seattle}, 2017, pp. 3483--3487.

\bibitem{ScaAst:15b}
G.~Scarciotti and A.~Astolfi, ``Model reduction of neutral linear and nonlinear
  time-invariant time-delay systems with discrete and distributed delays,''
  \emph{IEEE Transactions on Automatic Control}, vol.~61, no.~6, pp.
  1438--1451, 2016.

\bibitem{GugAntBea:08}
S.~Gugercin, A.~C. Antoulas, and C.~Beattie, ``$\mathcal{H}_2$ model reduction
  for large-scale linear dynamical systems,'' \emph{SIAM Journal on Matrix
  Analysis and Applications}, vol.~30, no.~2, pp. 609--638, 2008.

\bibitem{ScaAst:16a}
G.~Scarciotti and A.~Astolfi, ``Data-driven model reduction by moment matching
  for linear and nonlinear systems,'' \emph{Automatica}, vol.~79, pp. 340--351,
  May 2017.

\bibitem{FILE_MRSS}
G.~Scarciotti, ``Simulation resources: {O}n {M}oment {M}atching for
  {S}tochastic {S}ystems,''
  \url{https://www.imperial.ac.uk/people/g.scarciotti/research.html}, accessed:
  2020-11-04.

\bibitem{VaLPit:93}
C.~F. {Van Loan} and N.~Pitsianis, \emph{Approximation with Kronecker
  Products}.\hskip 1em plus 0.5em minus 0.4em\relax Dordrecht: Springer
  Netherlands, 1993, pp. 293--314.

\bibitem{Gen:07}
M.~G. Genton, ``Separable approximations of space-time covariance matrices,''
  \emph{Environmetrics}, vol.~18, no.~7, pp. 681--695, 2007.

\bibitem{ScaJiaAst:17}
G.~Scarciotti, Z.~P. Jiang, and A.~Astolfi, ``Data-driven constrained optimal
  model reduction,'' \emph{European Journal of Control}, vol.~53, pp. 68--78,
  May 2020.

\bibitem{ScaJiaAst:16}
------, ``Constrained optimal reduced-order models from input/output data,'' in
  \emph{\CDC{55th}, Las Vegas, NV, USA, December 12-14}, 2016, pp. 7453--7458.

\bibitem{RogWil:94}
L.~C.~G. Rogers and D.~Williams, \emph{Diffusions, {M}arkov Processes, and
  Martingales: {V}olume 1, {F}oundations}, ser. Cambridge Mathematical
  Library.\hskip 1em plus 0.5em minus 0.4em\relax Cambridge University Press,
  1994.

\bibitem{Koz:69}
F.~Kozin, ``A survey of stability of stochastic systems,'' \emph{Automatica},
  vol.~5, no.~1, pp. 95--112, 1969.

\bibitem{Kha:11}
R.~Khasminskii and G.~N. Milstein, \emph{Stochastic Stability of Differential
  Equations}, ser. Stochastic Modelling and Applied Probability.\hskip 1em plus
  0.5em minus 0.4em\relax Springer Berlin Heidelberg, 2011.

\end{thebibliography}

\vfill

\end{document}